\documentclass[useAMS,usenatbib,psfig]{mn2e} 
\usepackage{graphicx}
\title[Early-type galaxies at large galactocentric radii - I]{Early-type galaxies at large galactocentric radii - I. \\Stellar kinematics and photometric properties }

\author[Max Spolaor et al.]{Max Spolaor$^{1,2}$\thanks{AAO Research Fellow; ms@aao.gov.au}, George K. T. Hau$^1$, Duncan A. Forbes$^1$, Warrick J. Couch$^1$\\
$^1$Centre for Astrophysics \& Supercomputing, Swinburne University, Hawthorn, VIC 3122, Australia\\
$^2$Australian Astronomical Observatory, PO Box 296, Epping, NSW 1710, Australia\\}

\begin{document}
\date{Accepted... Received...; in original form 2010}

\pagerange{\pageref{firstpage}--\pageref{lastpage}} \pubyear{2010}

\maketitle
\label{firstpage}

\begin{abstract}
We present the results of a combined analysis of the kinematic and photometric properties at large galactocentric radii of a sample of 14 low-luminosity early-type galaxies in the Fornax and Virgo clusters. From Gemini South GMOS long-slit spectroscopic data we measure radial profiles of the kinematic parameters $v_{rot}$, $\sigma$, $h_{3}$, and $h_{4}$ out~to~$\sim 1 - 3$ effective radii. Multi-band imaging data from the HST/ACS are employed to evaluate surface brightness profiles and isophotal shape parameters of ellipticity, position angle and discyness/boxiness. The galaxies are found to host a cold and old stellar component which extend to the largest observed radii and that is the dominant source of their dynamical support. The prevalence of discy-shaped isophotes and the radial variation of their ellipticity are signatures of a gradual gas dissipation.
An early star-forming collapse appears to be the main mechanism acting in the formation of these objects. Major mergers are unlikely to have occurred in these galaxies. We can not rule out a minor merging origin for these galaxies, but a comparison of our results with model predictions of different merger categories places some constraints on the possible merger progenitors. These merger events are required to happen at high-redshift (i.e., $z \geq 1$), between progenitors of different mass ratio (at least 3:1) and containing a significant amount of gas (i.e., $\geq 10$ percent). A further scenario is that the low-luminosity galaxies were originally late-type galaxies, whose star formation has been truncated by removal of gas and subsequently the disc has been dynamically heated by high speed encounters in the cluster environment.

\end{abstract}

\begin{keywords} galaxies: dwarf $-$ galaxies: elliptical and lenticular, cD $-$ galaxies: formation $-$  galaxies: evolution $-$ galaxies: photometry $-$ galaxies: kinematics and dynamics
\end{keywords}

\section{Introduction}
In the past years, an increasing number of observational studies have focused their effort in probing kinematic, photometric and stellar population properties at large galactocentric radii of early-type galaxies (e.g., \citealt{hau06}; \citealt{spolaor09a}; \citealt{proctor09}; \citealt{weijmans09}; \citealt{foster09}).
This interest is motivated by the fact that nuclear regions are indicative of only $\sim$15 percent of the total stellar mass and of less than 10 percent of the angular momentum of a galaxy (i.e., assuming 
a de Vaucouleurs r$^{1/4}$ light profile and a flat rotation profile). 

In numerical simulations, the efficiency of galaxy formation mechanisms are predicted to vary with radius leaving measurable changes at different galactocentric radii.
The spatial distribution of these observed stellar properties, and their relationship to galaxy structural parameters, provide unique clues to constrain competing galaxy formation processes.

A specific example is the mass-metallicity gradient relation found by \cite{spolaor09a}. The correlation shows a sharp change in slope at a dynamical mass of  $\sim$ 3.5$\times 10^{10}$~M$_{\odot}$ (i.e., $M_{B} \sim -19$). The metallicity gradients of more massive galaxies (i.e., high-luminosity galaxies) flatten with increasing mass, perhaps due to the progressively larger number of mergers that these galaxies undergo to assemble their mass. In the remnants, the gradients are partially regenerated if a central burst of star formation is induced by the merger. In low-mass (i.e., low-luminosity) galaxies, an early star-forming collapse with a varying star formation efficiency is required in order to explain the steepening of gradients with increasing mass. The depth of the galaxy gravitational potential well coupled with energy feedback via supernova-driven galactic winds are responsible for shaping the gradients.

The effect of gas dissipation, merger-induced starbursts, and energy feedback are crucial ingredients in galaxy formation models to be able to reproduce the observed stellar properties of early-type galaxies (e.g., \citealt{kawata03}; \citealt{pipino06}; \citealt{springel05}; \citealt{robertson06}). Specifically, the interplay and time ordering between dissipative processes and the redistribution of angular momentum could be the primary sources of the different stellar properties observed in low-luminosity and high-luminosity galaxies (e.g., \citealt{rix99}; \citealt{dimatteo09}; \citealt{hoffman09}). 

The variety of kinematic and photometric features displayed by early-type galaxies, such as isophotal shapes, extended stellar discs, amount of ordered and random motion are interpreted as signatures of these mechanisms. For example, the results of the SAURON survey (e.g., \citealt{bacon01}) have prompted the division of early-type galaxies into slow and fast rotators, depending on the amount of stellar angular momentum per unit mass inside one effective radius (\citealt{emsellem07}; \citealt{krajnovic08}). 

The aim of this paper is to complement the stellar populations findings of \cite{spolaor09c} (hereafter Paper II) on galaxy formation mechanisms acting in low-luminosity early-type galaxies.
We do this by exploring the kinematic and photometric properties at large galactocentric radii of the same 14 low-luminosity galaxies.

In general, properties of low-luminosity early-type galaxies have not been fully probed to large radii (e.g., \citealt{davies83}; \citealt{bender94}; \citealt{halliday01}). 
The low-surface brightness of these objects makes the measurement of reliable properties at large radii observationally challenging.
Here, we study the galaxy kinematics by analysing the line-of-sight velocity distribution (LOSVD) as a function of galactocentric radius from high quality Gemini GMOS long-slit spectroscopic data.
We derive spatially resolved radial profiles out to $\sim 1 - 3$ effective radii for rotation velocity, velocity dispersion and the Gauss-Hermite coefficients $h_{3}$ and $h_{4}$. The higher order moments of the distribution allow us to derive symmetric and asymmetric deviations from a pure Gaussian shape. They relate to the amount of ordered motion and the shape of stellar orbits (e.g. \citealt{marel93}; \citealt{bender94}; \citealt{cappellari04}). The isophotal properties of a galaxy, such as surface brightness profile, ellipticity, position angle and discyness/boxiness are obtained from high-resolution HST/ACS imaging data in the $g$ and $z$ filters.

The paper is organised as follows. In Section 2, we describe the data sample. In Section 3, we describe observations and reduction of spectroscopic and imaging data.
Section 4 the technique used to recover stellar kinematics is described and the results are presented. In Section 5, we describe the surface photometry analysis and the results are presented. In Section 6, the combined kinematic and photometric properties of the galaxies are discussed. In Section 7, we discuss our results in the context of predictions from competing galaxy formation scenarios. In Section 8, conclusions are given.

\section{Data sample} 
The data sample consists of 14 low-mass early-type galaxies in the Fornax and Virgo clusters. 
The six galaxies in the Fornax cluster are chosen from the catalogue of \cite{ferguson89}, while the eight galaxies in the Virgo cluster are from the catalogue of \cite{binggeli85}.
Galaxies of low mass are selected on the basis on their central stellar velocity dispersion $\sigma_{0}$ and the absolute $B$-band total magnitude $M_{B}$, both of which are known to be independent proxies of galaxy mass. 
The sample uniformly covers the range $1.6 < \log(\sigma_{0}) < 2.15$ and $-16.5 > M_{B} > -19.5$. 
This translates to a dynamical mass range of about $10^{9} <~$M$_{\rm dyn}/\rm M_{\odot} < 10^{11}$, using the $\log($M$_{\rm dyn}) = 2 \log(\sigma_{0}) + \log(r_{e})+ 5.0$ where $r_{e}$ is the effective radius in parsecs (\citealt{cappellari06}).
The main properties of the galaxy sample are summarised in Table~\ref{gal_prop}.

\begin{table*}
\begin{center}
\begin{tabular}{cccccccccccc}
\hline 
\hline 
Galaxy & Alternative & Hubble & $r_{e}$  & P.A.&  Distance & $M_{B}$ & M$_{\rm dyn}$ & Radial Range\\
            &     Name    &  Type  & [arcsec] & [degree]&   [Mpc] &[mag] &[10$^{9}$~M$_{\odot}$] & $r_{e}$ (kpc)\\ 
(1)       &  (2)            & (3) & (4) &(5) &(6) & (7) &(8) & (9)  \\
\hline 
FCC~148 & NGC~1375 & S0 & 14.9  &  89 & 18.45 &   -17.30 & 7.9 & 2.0 (2.66)\\
FCC~153 & IC~1963     & S0 & 12.4  &  81 &  18.45 & -17.10 & 3.1 & 3.4 (4.02)\\
FCC~170 & NGC~1381 & S0 & 13.7 &   139 &  18.45 & -18.88 & 34.6& 3.4 (4.76)\\
FCC~277 & NGC~1428 & E/S0 & 9.8 &   123 &   18.45 & -17.52 & 9.5 &2.0 (1.90)\\
FCC~301 & ESO~358-G059 & E/S0 & 10.4 & 156 &  18.45 & -17.02 & 3.9 & 1.0 (0.58)\\
FCC~335 & ESO~359-G002 & E/S0 & 15.6 & 47 & 18.45 & -16.93 & 3.5 & 1.0 (1.40)\\
VCC~575 & NGC~4318 & E4 & 7.2 &  65  &    22.08 &  -17.61 & 7.2 & 2.1 (1.62) \\
VCC~828 & NGC~4387 & E5 & 9.1 &   137  &  17.94 &  -18.26 & 10.7 & 2.0 (1.58) \\
VCC~1025 & NGC~4434 & E0/S0 & 12.9 & 134 & 22.39 &  -18.72 & 31.6 & 2.0 (2.80)\\
VCC~1146 & NGC~4458 & E1 & 26.1 & 175  &  16.37 & -18.13 & 26.9 & 0.8 (1.66)\\
VCC~1178 & NGC~4464 & E3 & 6.6 &  14  & 	15.84 &  -17.63 & 7.2 & 3.2 (1.62)\\
VCC~1297 & NGC~4486B & E1 & 2.3 &  110 &  16.29 &  -16.77 & 16.2 & 3.0 (0.55)\\
VCC~1475 & NGC~4515 & E2 & 9.4 &   15  &  16.59 &  -17.84 & 7.7&2.0 (1.51)\\
VCC~1630 & NGC~4551 & E2 & 12.5 &  67  &  16.14 &  -18.06 & 1.0 &2.2 (2.15)\\
\hline \hline
\end{tabular}
\end{center}
\caption[]{ Galaxy properties. Col.(1): galaxy name from the catalogue
 of \cite{ferguson89} and \cite{binggeli85}; (2): alternative galaxy name; (3): morphological type from the HyperLEDA database; (4): $B$-band effective radius along the galaxy semi-major axis from the HyperLEDA database; (5): position angle of the major axis; (6): distance to the galaxy using the distance modulus derived from surface brightness fluctuations of \cite{tonry01}, \cite{mei05} and \cite{blakeslee09}; we applied the correction of \cite{jensen03} to the values of Tonry; (7): total $B$-band absolute magnitudes estimated from total apparent magnitudes given in the NASA/IPAC Extra-galactic Database and corrected for Galactic extinction; (8): dynamical mass of the galaxy using the $\log($M$_{\rm dyn}) = 2 \log(\sigma_{0}) + \log(r_{e})+ 5.0$, where $r_{e}$ is the effective radius expressed in parsecs (\citealt{cappellari06}); (9): radial range in effective radius and kpc (in parenthesis) covered by our observations.}
\label{gal_prop}
\end{table*}

\section{Observations and Data reduction}
\subsection{GMOS long-slit spectroscopy}
The observations were made using Gemini Multi-Object Spectrograph (GMOS) mounted on the Gemini South telescope in long-slit spectroscopy mode. The six galaxies in the Fornax cluster were observed during the semester 2006B (Program ID GS-2006B-Q-74), while the eight galaxies in the Virgo cluster were observed during the semester 2008A (Program ID GS-2008A-Q-3). 

Four 1800s exposure were taken for each galaxy except for FCC~335, for which two exposures of 1800s were taken.  At the beginning of each exposure, the slit was centred on the nucleus and oriented along the major axis of the galaxy. The GMOS detector comprises three 2048~$\times$~4608~EEV~CCDs with a pixel size of 13.5~$\times$13.5~$\mu$m$^{2}$. We adopted the standard full frame readout, and to enhance the signal-to-noise, we selected to bin 2$\times$~spatially, yielding a spatial resolution of 0.1454~arcsec~pixel$^{-1}$. 

The spectrograph slit width was set to 1.0~arcsec for the 2006B run and 0.5~arcsec for the 2008A run. It was 5.5$\arcmin$ in length in both cases. For the 2006B run, we used the 600 line~mm$^{-1}$ grism blazed at 5075~\AA,~providing a dispersion of 0.457~\AA~pixel$^{-1}$ and a wavelength coverage of 3670$-$6500~\AA. For the 2008A run, we used the 600 line~mm$^{-1}$ grism blazed at 5250~\AA~with 0.458~\AA~pixel$^{-1}$, covering the wavelength range of 3860$-$6670~\AA. The instrumental set up provided a spectral resolution, as measured from the full width at half maximum~(FWHM) of the arc lines, of $\sim$~4.21~\AA~(2006B run) and $\sim$~2.45~\AA~(2008A run). During the observations, the seeing varied from 0.7-arcsec to 1.0-arcsec FWHM, which is more than adequate for the purposes of our study. Spectrophotometric and radial velocity standard stars covering a range of spectral classes were observed as calibrators. These stars belong to both the Lick/IDS stellar library (\citealt{worthey94}) and MILES library (\citealt{patrizia06}). Bias frames, dome flat-fields and copper$-$argon (CuAr) arcs were also taken for calibration. 

The data reduction was carried out using the Gemini/GMOS IRAF package (version 1.9.1). Each object frame is treated separately. Initial reduction of the CCD frames involves overscan correction, bias and dark subtraction, flatfield correction and cosmic ray removal. The flatfield correction is performed by means of both a quartz-halogen lamp (taken with the Gemini calibration unit, GCAL) and twilight spectra, which are normalised and divided from all the spectra. This process corrects for pixel-to-pixel sensitivity variations and for two-dimensional large-scale illumination patterns due to slit vignetting. Cosmic rays are identified and eliminated by interpolating each line of an image with a high order function. The residuals (cosmic ray hits) that are narrower than the expect instrumental line width are replaced with the fitted value. The package GSWAVELENGTH is used to establish an accurate wavelength calibration from copper$-$argon (CuAr) arc images taken just before and after each exposure. Each spectrum is converted to a linear wavelength scale using roughly 130 arc lines fitted by 3rd-5th order polynomials. An accuracy of better than 0.1~\AA~is consistently achieved for the wavelength calibration solution along the entire spectral coverage. The spectra are also corrected for geometrical distortions along the spectral directions (S-distortion). The spectral resolution is derived as the mean of the Gaussian FWHMs measured for a number of unblended arc-lamp lines which are distributed over the whole spectral range of a wavelength-calibrated spectrum. 

To obtain reliable measurements of the kinematic parameters at large galactocentric radii, it is critical to perform an accurate sky subtraction. Sky contamination becomes increasingly significant for the outer parts of the galaxy. In these regions, the level of galaxy light can be a few percent of the sky signal. This effect represents one of the main sources of systematic errors in the analysis of the kinematic radial profiles. The generous field-of-view of GMOS and the intrinsic low luminosity of our sample allow us to accurately estimate the sky continuum from the edges of the slit, where the galaxy light is negligible. The sky level is estimated by interpolating along the outermost 10$-$20 arcsec at the two edges of the slit and subtracted from the whole image. A sky subtraction accurate to within one percent is achieved, guaranteeing negligible related systematics on the measured kinematic parameters.

For each galaxy, we obtained four fully reduced galaxy frames of identical exposure time. These are then co-added to form a single frame. The final galaxy frame is a two-dimensional spectrum from which we extract 1D spectra along the slit. The spatial width (i.e.~the number of CCD rows binned) for each extracted 1D spectrum increases with radius to achieve a minimum signal-to-noise (S/N) per \AA~of $\sim$~35 in the spectral region of Mg$_{b}$. By adopting this criterion we have been able to obtain reliable measurements of kinematic parameters to galactocentric radii of $\sim$~3$r_{e}$.

\subsection{HST/ACS imaging}
The imaging observations were performed with the Advanced Camera for Surveys (ACS) mounted on the Hubble Space Telescope (HST).
The eight Virgo cluster galaxies were observed by the ACS Virgo Cluster Survey (ACSVCS; \citealt{cote04}), while the six galaxies in the Fornax cluster were observed by the ACS Fornax Cluster Survey (ACSFCS; \citealt{jordan07}).
Each galaxy was observed using the ACS Wide Field Channel (WFC) mode (\citealt{ford98}; \citealt{sirianni05}).
The WFC consists of two 2048$\times$4096 CCD detectors (WFC1 and WFC2) with a pixel size of 15$\mu$m and a spatial scale of 0.049$\arcsec$ pixel$^{-1}$.
The nucleus of each galaxy was placed on the WFC1 detector at the reference pixel (2096, 200). To avoid the gap between the WFC1 and WFC2 detectors an additional offset 5$\arcsec$ to 10$\arcsec$ was applied (depending by the brightness of the target galaxy). 

Each galaxy was observed with the same observational methodology. Two images in the F475W filter (which resembles the SDSS $g$ filter) for a total of 760s, with a respective exposure time of 380s to aid the removal of cosmic rays. Two images in the F850LP filter (similar to the SDSS $z$ filter) for a total of 1220s, split in a pair of 760s exposures. A further image of 90s exposure in F850LP filter was taken to recover the central regions of the galaxies whose nuclei might saturate in the longer 560s exposure.

The data were downloaded from the Multimission Archive at the Space Telescope Science Institute (MAST).
The images were processed by the ACS calibration pipeline (CALACS; \citealt{pavlo06}) using the ``On the Fly Reprocessing'' option.
An automated data reduction pipeline was developed for the ACSVCS data (\citealt{jordan04}).
The data reductions procedures for the ACSFCS data have been documented in \cite{jordan07} and they are similar to those adopted for the ACSVCS data.
Briefly, CALACS performed bias and dark subtraction, flat-fielding, and overscan removal.
The five images of each galaxy were aligned by matching coordinates for a list of objects in common, and images taken with the same filter were then combined.
A cosmic rays rejection procedure was applied before performing geometric distortion correction to combined images.
The field-of-view of the fully reduced images is  approximately 202$\arcsec$$\times$202$\arcsec$ in the shape of a rhomboid.

\section{Galaxy Stellar Kinematics}
We use the penalised pixel fitting (pPXF) code of \cite{cappellari04} to extract kinematic parameters of the galaxies.
This is based on the method for identification of non-Gaussian line profiles, originally developed by \cite{marel93}.

In the pPXF software the line-of-sight velocity distribution (LOSVD) is parametrized as a sum of orthogonal functions in a Gauss-Hermite series, namely a Gaussian plus a series of Gauss-Hermite polynomials.
A $\chi^{2}$ minimising process determines the best-fitting moments of the LOSVD.
The mean line-of-sight velocity, $v_{\rm los}$, the velocity dispersion, $\sigma$, and the coefficients of the Gauss-Hermite polynomials of third, $h_{3}$, and forth, $h_{4}$, order.

The coefficients $h_{3}$ and $h_{4}$ quantify asymmetric (skewness) and symmetric (kurtosis) deviations of the LOSVD from a simple Gaussian distribution, respectively.
Positive $h_{3}$ values indicate a distribution skewed towards velocities lower than the systemic velocity of the galaxy, $v_{sys}$.
Conversely, negative values of $h_{3}$ are characteristic of a distribution skewed toward velocities greater than $v_{sys}$. 
Positive $h_{4}$ values correspond to a distribution more peaked than a Gaussian, while negative values of $h_{4}$ denote a flatter distribution.
 
The software works in pixel space, finding the linear combination of template stars spectra which, after being convolved with an appropriate LOSVD, best reproduces the galaxy spectrum.
The advantage of this technique is that errors due to template-mismatch are reduced by using a combination of template stars.
In our analysis, ten template stars of spectral type in the range from G3 to K9 are used.
The entire wavelength range covered by the galaxy spectrum is modelled, but the code allows us to mask out regions possible affected by skyline residuals or emission lines.
A pixel-by-pixel $\chi^{2}$ minimisation of the residuals to the best fit is carried out. An adjustable penalty term is applied to the $\chi^{2}$ to bias the higher order moments coefficients $h_{3}$ and $h_{4}$ of the 
resulting LOSVD towards a Gaussian shape. We find that a small bias of 0.4 provides an optimum value for the spectra of our data sample.

We estimate random errors on the derived parameters using a series of Monte Carlo simulations. 
The entire kinematic measurement process is repeated on a large number of different realisations of our data sample. 
The model spectra are obtained using the corresponding best linear combination of template stars, broadened by the best-fitting LOSVDs and degraded to a range of S/N values.

\subsection{Radial profiles}
\begin{figure*}
\includegraphics[scale=.7]{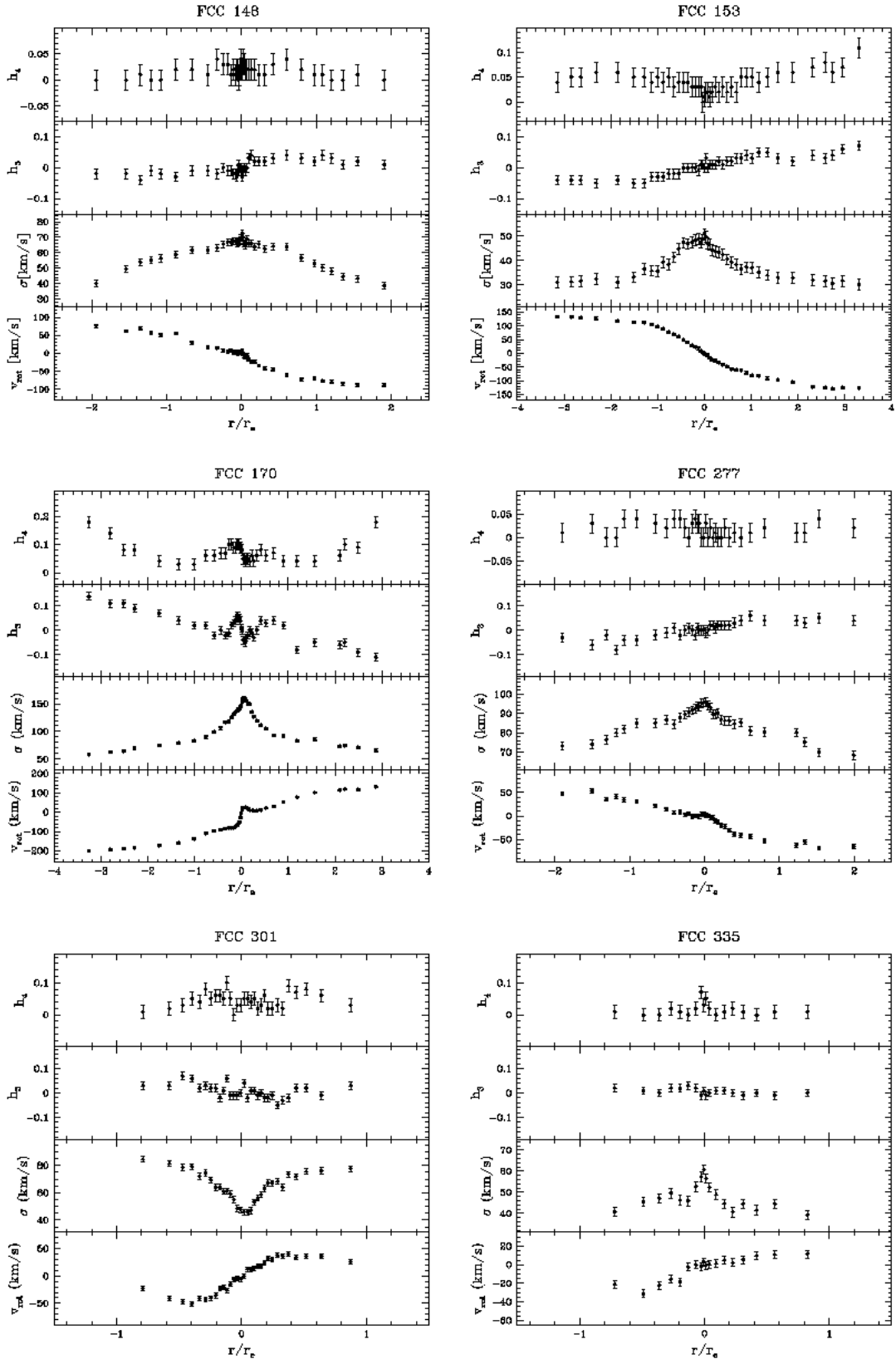}
\caption[]{The major-axis stellar kinematics of the Fornax cluster low-luminosity galaxies. From bottom to top: radial profiles of rotational velocity $v_{rot}$, velocity dispersion $\sigma$ and the Gauss-Hermite coefficients $h_{3}$ and $h_{4}$. The kinematic quantities are expressed as a function of the galactocentric radius of each spatially resolved aperture, normalised with the effective radius of the galaxy's semi-major axis.}
\label{kin_fornax}
\end{figure*}

\begin{figure*}
\includegraphics[scale=.7]{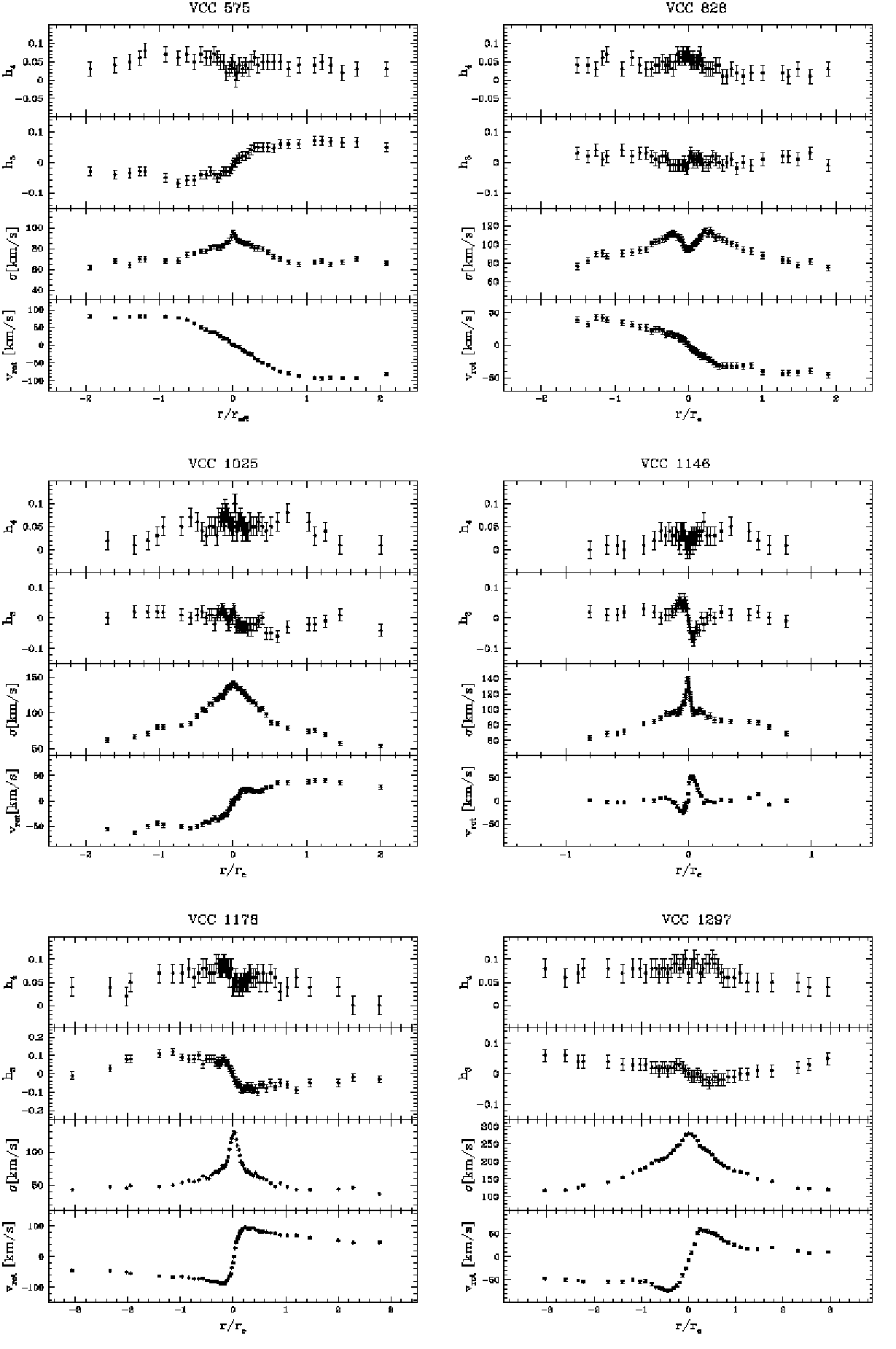}
\caption[]{The major-axis stellar kinematics of the Virgo cluster low-luminosity galaxies. From bottom to top: radial profiles of rotational velocity $v_{rot}$, velocity dispersion $\sigma$ and the Gauss-Hermite coefficients $h_{3}$ and $h_{4}$. The kinematic quantities are expressed as a function of the galactocentric radius of each spatially resolved aperture, normalised with the effective radius of the galaxy's semi-major axis.}
\label{kin_virgo1}
\end{figure*}

\begin{figure*}
\includegraphics[scale=.7]{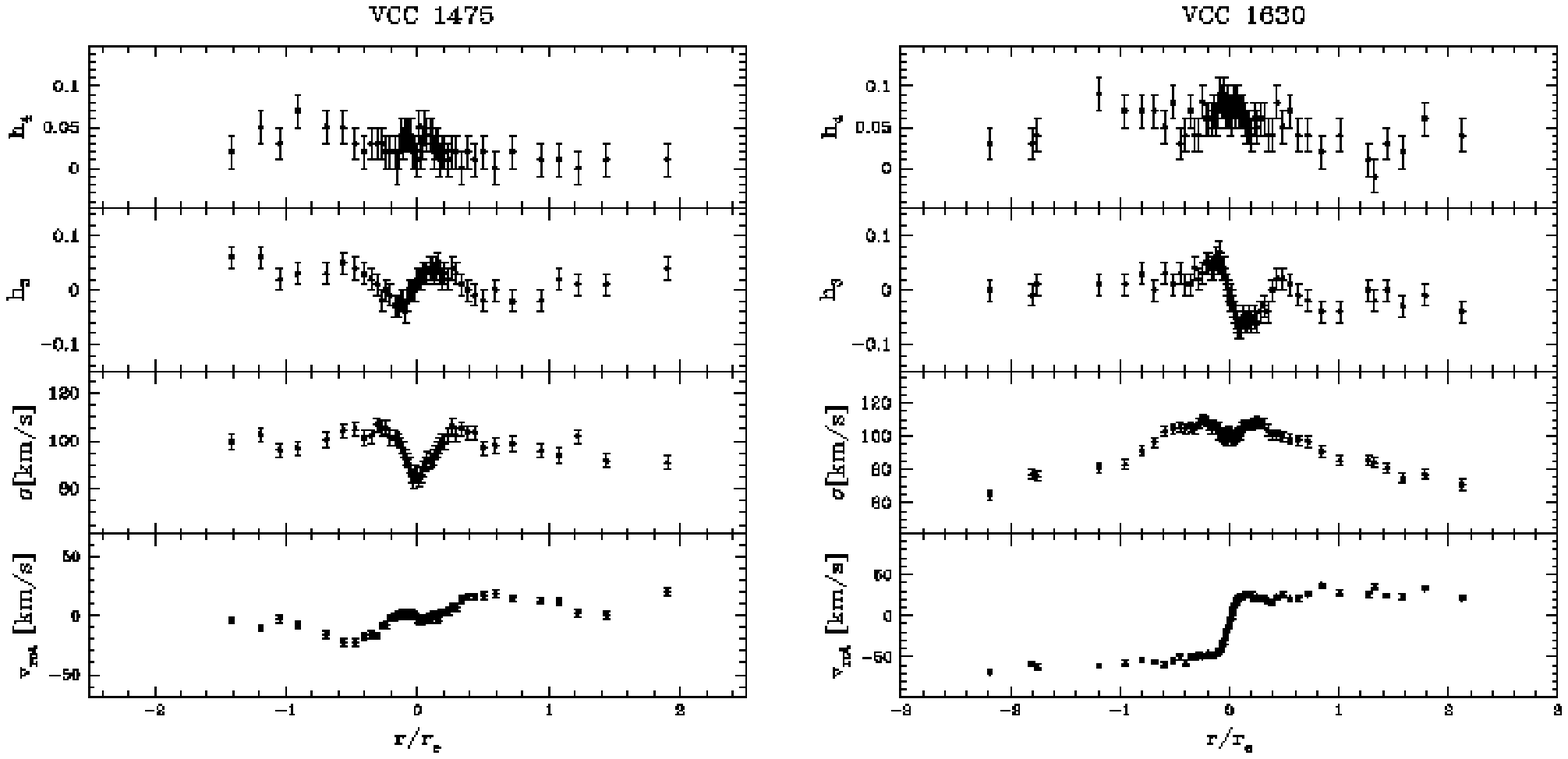}
\caption[]{The major-axis stellar kinematics of the Virgo cluster low-luminosity galaxies.  Continued from Fig.~\ref{kin_virgo1}.}
\label{kin_virgo2}
\end{figure*}

The radial profiles of the kinematic parameters $v_{\rm rot}$, $\sigma$, $h_{3}$, and $h_{4}$ for the 14 low-mass galaxies are presented in Fig.~\ref{kin_fornax}, Fig.~\ref{kin_virgo1}, and Fig.~\ref{kin_virgo2}. 
The parameters are expressed as a function of the galactocentric radius along the galaxy's major-axis of each spatially resolved aperture, scaled with the effective radius of the galaxy's semi-major axis.
The rotation velocity $v_{\rm rot}$ is calculated as the difference between the measured $v_{\rm los}$ of an aperture and the $v_{\rm sys}$ of the galaxy, that is the median value of $v_{\rm los}$ from all apertures extracted along the slit for each galaxy.
A description of the individual galaxies is given in Appendix~\ref{ind_gals}.

To be able to compare our results with previous studies we derive the following parameters describing the kinematic properties of the galaxies.
The maximum measured rotation velocity $v_{\rm max}$ within a radius $r \leq r_{e}/2$.
The central $\sigma_{0}$ and mean $\overline{\sigma}$ velocity dispersion, obtained by averaging the $\sigma$ values estimated in all spatially resolved apertures within a radius $r \leq r_{e}/8$ and $r \leq r_{e}/2$, respectively. 
The mean of absolute values of $h_{3}$,  $\mid\overline{h_{3}}\mid$, and the mean of values of $h_{4}$, $\overline{h_{4}}$, estimated from all spatially resolved apertures within a radius $r \leq r_{e}/2$. The parameters for each galaxy are listed in Table~\ref{glob_pars}.

\begin{table*}
\begin{center}
\begin{tabular}{ccccccccccc}
\hline 
\hline 
Galaxy & $v_{\rm sys}$ & $v_{\rm max}$ & $\sigma_{0}$ & $\overline{\sigma}$ &  $\mid\overline{h_{3}}\mid$ & $\overline{h_{4}}$ & $\overline{g - z}$& $\overline{\epsilon}$ & $\overline{P.A.}$ & $\overline{B_{4}}$\\
            & [km s$^{-1}$] & [km s$^{-1}$] & [km s$^{-1}$] & [km s$^{-1}$] & & &&&[$^{\circ}$]&\\
(1)       &  (2)            & (3) & (4) &(5) &(6) & (7)&(8)&(9)&(10)& (11) \\            
\hline 
FCC~148 & 680.1&89.0&67.8&66.1&0.031&0.021&1.04&0.589&84.2&-0.0361 \\
FCC~153 & 1603.0&132.9&48.4&43.4&0.016&0.032&1.31& 0.612&75.1&0.0028	\\
FCC~170 & 1677.4&200.9&151.5&131.1&0.027&0.068&1.47& 0.480&137.7&-0.0100	\\
FCC~277 & 1595.0&68.2&93.7&89.0&0.017&0.016&1.37& 0.368&116.3&-0.0079	\\
FCC~301 & 1023.4&50.9&58.9&71.8&0.088&0.044	&1.25 & 0.440&149.2&0.0004	\\
FCC~335 & 1467.3&31.2&44.5&38.7&0.006&0.005	&1.15& 0.398&135.4&0.0014	\\
VCC~575 & 1168.3 & 92.3 &88.2 &80.5 &0.037 & 0.046&1.34&0.387&130.2& 0.0071	\\
VCC~828 & 543.6 &46.4&104.1&102.6&0.012&0.044&1.47& 0.378&36.2&	-0.0244\\
VCC~1025 &994.4 &62.2&135.0&120.6&0.020&0.057	&1.52& 0.050&34.5&0.0014\\
VCC~1146 &580.5 &53.4&102.8&97.7&0.029&0.025	&1.33 &  0.089&81.6&0.0034\\
VCC~1178 &1265.6 &94.7&108.2&77.7&0.070&0.065	&1.32 &  0.302&76.2&0.0071\\
VCC~1297 &1485.3&74.0&272.7&223.2&0.057&0.080	&1.50 &  0.144&22.4&0.0058\\
VCC~1475 &927.6 &22.4&91.8&98.8&0.023&0.027		&1.34 &  0.178&55.4&0.0076\\
VCC~1630 & 1187.9&68.4&109.0&109.9&0.035&0.063	&1.32 & 0.288&148.0& -0.0107\\
\hline \hline
\end{tabular}
\end{center}
\caption[]{Kinematic and photometric parameters for the 14 low-luminosity early-type galaxies. See text for details.}
\label{glob_pars}
\end{table*}

\section{Galaxy Surface Photometry}
The isophotal analysis of images in both filters of each galaxy is conducted using the IRAF task ELLIPSE.
The package is based on the method of \cite{jer87}, and it involves the creation of a smooth elliptical galaxy model to fit to the real galaxy image (e.g., \citealt{ferrarese06}).

The technique makes the assumption that each isophote can be modelled by an ellipse whose centre, ellipticity and position angle are allowed to vary to reach the best-fitting values. The model is built up from a first guess elliptical isophote whose centre is pre-determined. The semi-major axis of each isophote is then incremented logarithmically from the previously fitted ellipse. The process is iterated until specific convergency criteria are achieved. A standard k-sigma clipping routine is used to evaluate and to
flag divergent points, thus increasing the stability of the convergency process. A mask is created to exclude from the fit bright globular clusters, foreground stars, background galaxies and the gap between the individual CCD detectors.

The background-subtraction is based on the average of median values from several 10$\times$10 pixel boxes sampled in regions not affected by galaxy light. The conversion from the HST/ACS instrumental F475W and F850W filters system to the AB photometric system is performed using the synthetic coefficients from \cite{sirianni05}. In particular, the F475W and F850W filters are transformed into $g_{AB}$ and $z_{AB}$, respectively. 
Final magnitudes were corrected for Galactic extinction using the reddening values from the DIRBE dust maps of \cite{sch98}.

The errors in magnitude are computed from the rms scatter of intensity data along the fitted ellipse and the errors in the background subtraction. Errors in the Galactic extinction correction values and in the zero-points are negligible with respect to the errors in the intensity profiles. Ellipse geometry parameter errors are obtained from the internal errors in the harmonic fit. Harmonic amplitude errors are obtained from the fit errors after removal of all harmonics up to and including, the one being considered. 

\subsection{Radial profiles}
\begin{figure*}
\includegraphics[scale=.85]{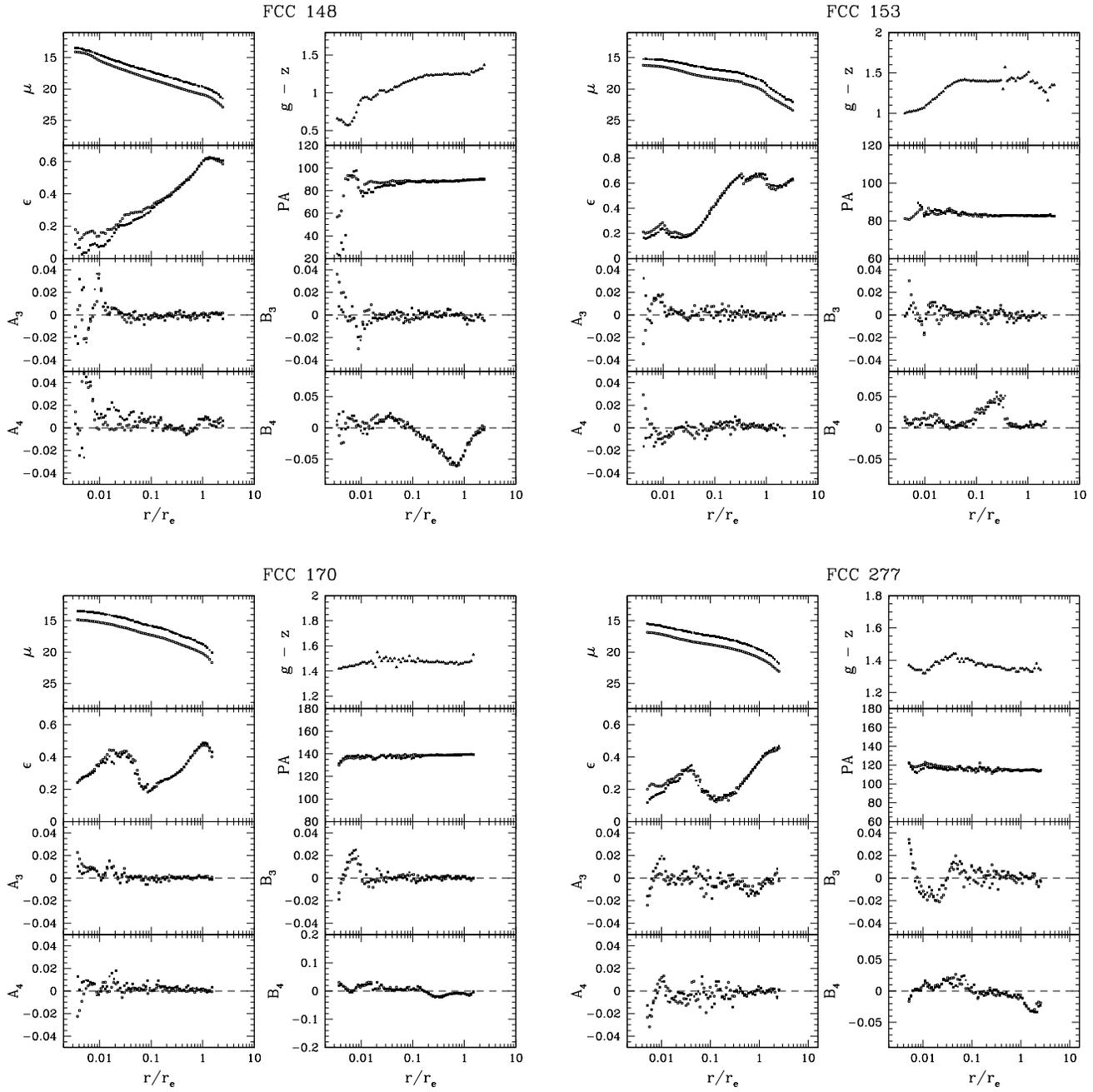}
\caption[]{Isophotal properties of the Fornax cluster low-luminosity galaxies. Open black squares indicate $g$-band parameters. Filled black squares denote $z$-band parameters.
In each plot are shown: surface brightness profiles $\mu$ and colour index profiles; ellipticity $\epsilon$, position angle $P.A.$ and best fit parameters $A_{3}$, $A_{4}$, and $B_{3}$, $B_{4}$.
Positive values of $B_{4}$ indicate discy isophotes, whereas negative values represent boxy isophotes. The photometric quantities are expressed as a function of the galactocentric radius of data points, normalised with the effective radius of the galaxy's semi-major axis.}
\label{phot_fornax1}
\end{figure*}

\begin{figure*}
\includegraphics[scale=.85]{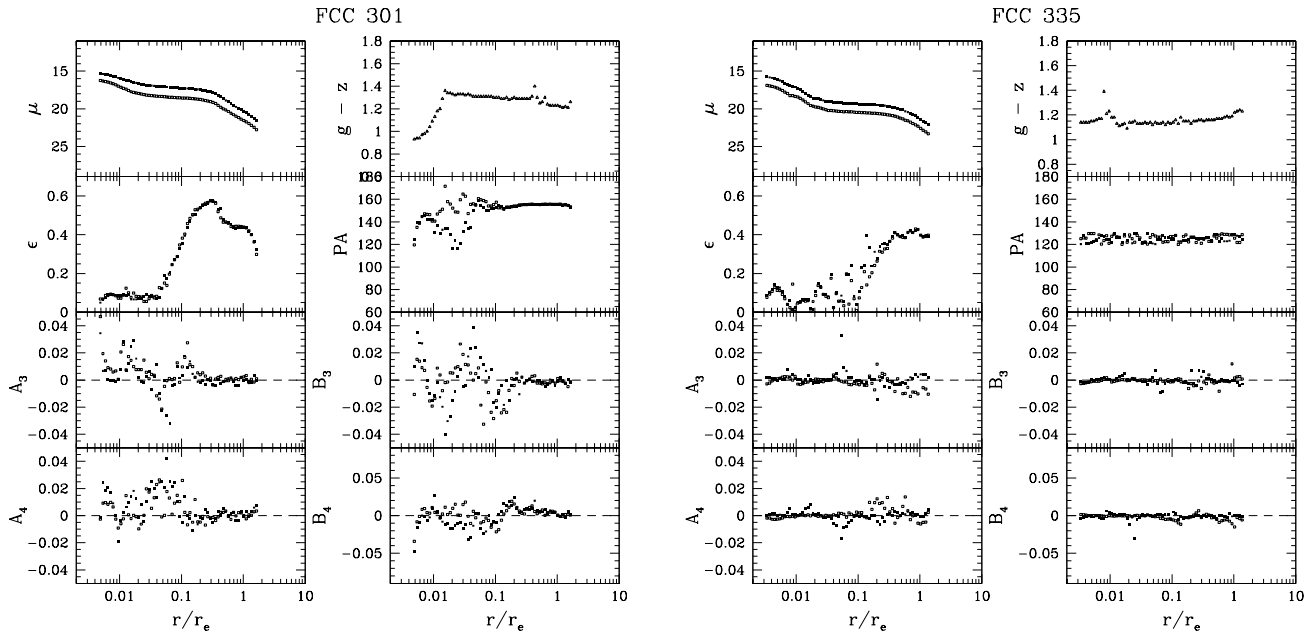}
\caption[]{Isophotal properties of the Fornax cluster low-luminosity galaxies.  Continued from Fig.~\ref{phot_fornax1}.}
\label{phot_fornax2}
\end{figure*}

\begin{figure*}
\includegraphics[scale=.85]{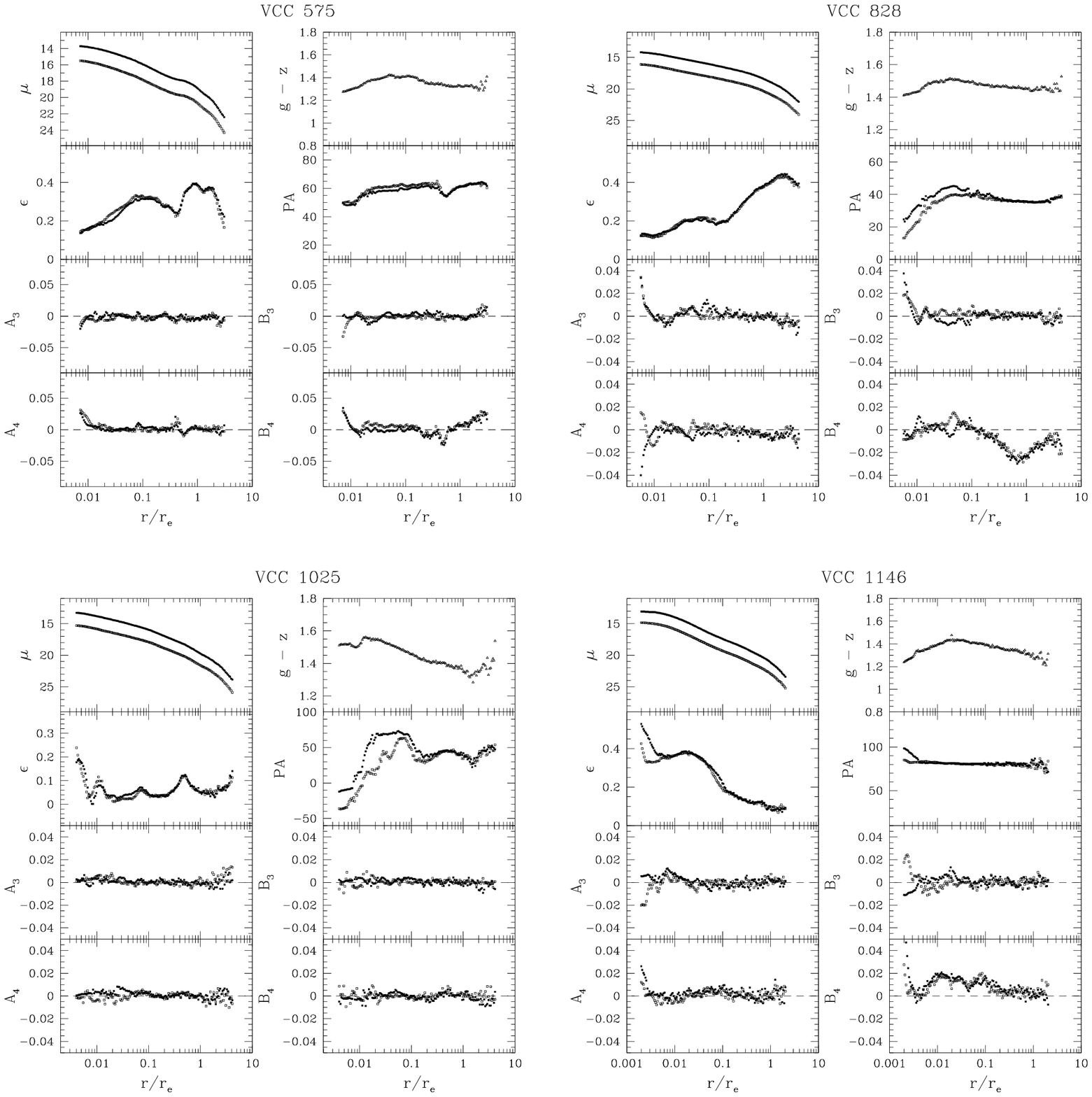}
\caption[]{Isophotal properties of the Virgo cluster low-luminosity galaxies. Open black squares indicate $g$-band parameters. Filled black squares denote $z$-band parameters.
In each plot are shown: surface brightness profiles $\mu$ and colour index profiles; ellipticity $\epsilon$, position angle $P.A.$ and best fit parameters $A_{3}$, $A_{4}$, and $B_{3}$, $B_{4}$.
Positive values of $B_{4}$ indicate discy isophotes, whereas negative values represent boxy isophotes. The photometric quantities are expressed as a function of the galactocentric radius of data points, normalised with the effective radius of the galaxy's semi-major axis.}
\label{phot_virgo1}
\end{figure*}

\begin{figure*}
\includegraphics[scale=.85]{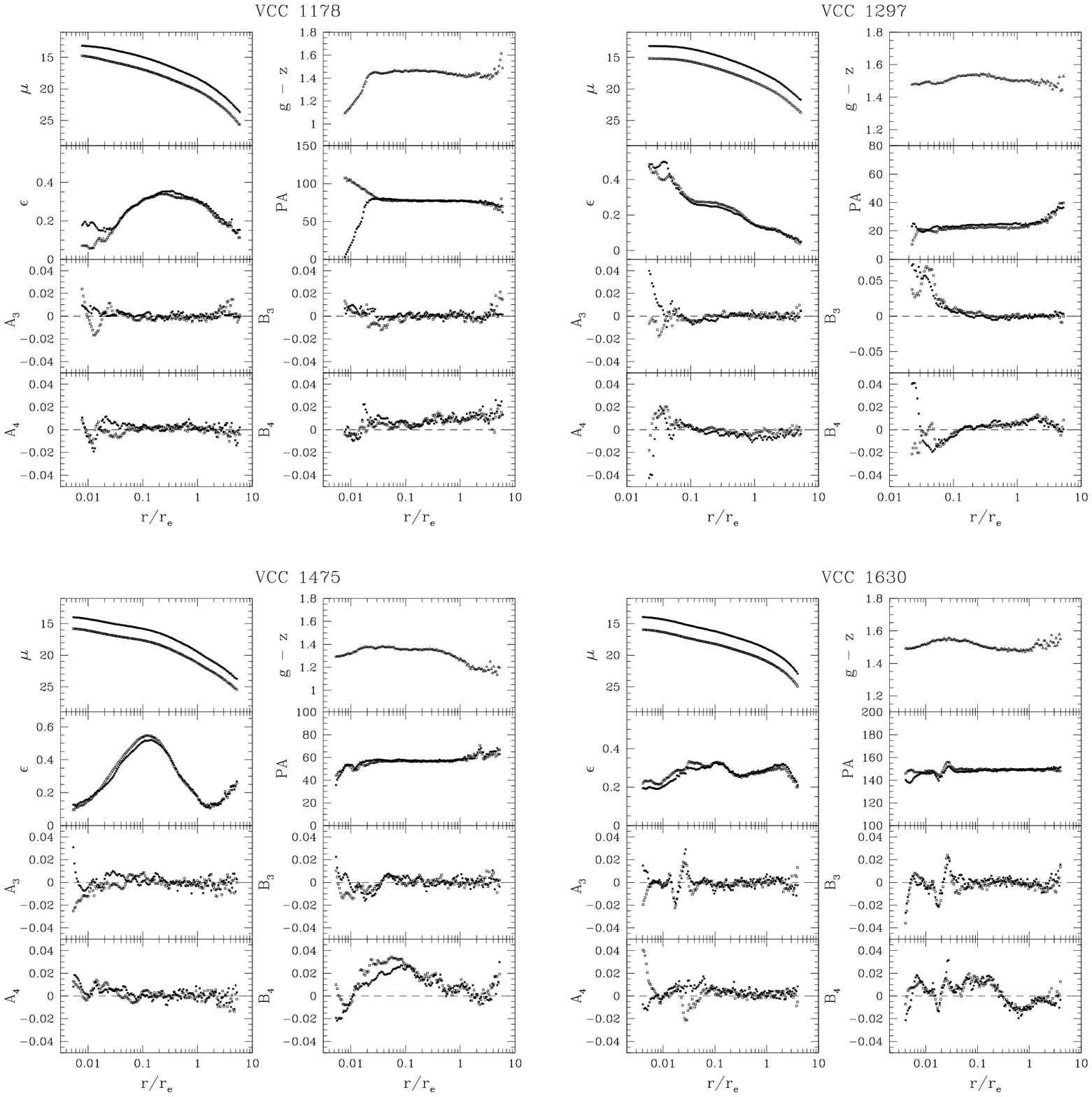}
\caption[]{Isophotal properties of the Virgo cluster low-luminosity galaxies.  Continued from Fig.~\ref{phot_virgo1}.}
\label{phot_virgo2}
\end{figure*}

We present the results of the isophotal analysis for the 14 sample galaxies in Fig.~\ref{phot_fornax1}, Fig.~\ref{phot_fornax2}, Fig.~\ref{phot_virgo1}, and Fig.~\ref{phot_virgo2}.
Photometric parameters are plotted as function of the geometric mean radius $r$, defined as $r=a[1-\epsilon(a)]^{1/2}$ where $a$ is the length of the isophotal semi-major axis. 
This radius is then scaled with the galaxy's effective radius along the semi-major axis and expressed as $r/r_{e}$.

For each galaxy, the $\mu_{g}$, $\mu_{z}$ surface brightness profiles and the $g$~-~$z$ colour index profile are shown.
We also plot radial profiles of isophotal geometry parameters of ellipticity $\epsilon$, position angle $P.A.$ and the isophote's deviations from perfect ellipticity $A_{3}$, $A_{4}$, and $B_{3}$, $B_{4}$.
These parameters are respectively the third and fourth cosine and sine terms of a Fourier expansion used by the software.
The coefficients $A_{3}$ and $B_{3}$ indicate isophotes with three-fold deviations from ellipses, i.e.~egg-shaped or heart-shaped.
Four-fold deviations are represented by the $A_{4}$ and $B_{4}$ coefficients, such that a rhomboidal or diamond shaped isophote translates into a nonzero $A_{4}$ value.
A positive value for $B_{4}$ indicates discy isophotes, whereas a negative value represents boxy isophotes.

Photometric parameters are derived by averaging all values measured inside half the effective radius in both filters.
Mean parameter values of colour index $\overline{g - z}$, ellipticity $\overline{\epsilon}$, position angle $\overline{P.A.}$, coefficient $\overline{B_{4}}$ for all 14 galaxies are listed in Table~\ref{glob_pars}.

\section{Galaxy properties}
\subsection{$v_{rot}/ \sigma$ radial profiles}
\begin{figure}
\includegraphics[scale=.4]{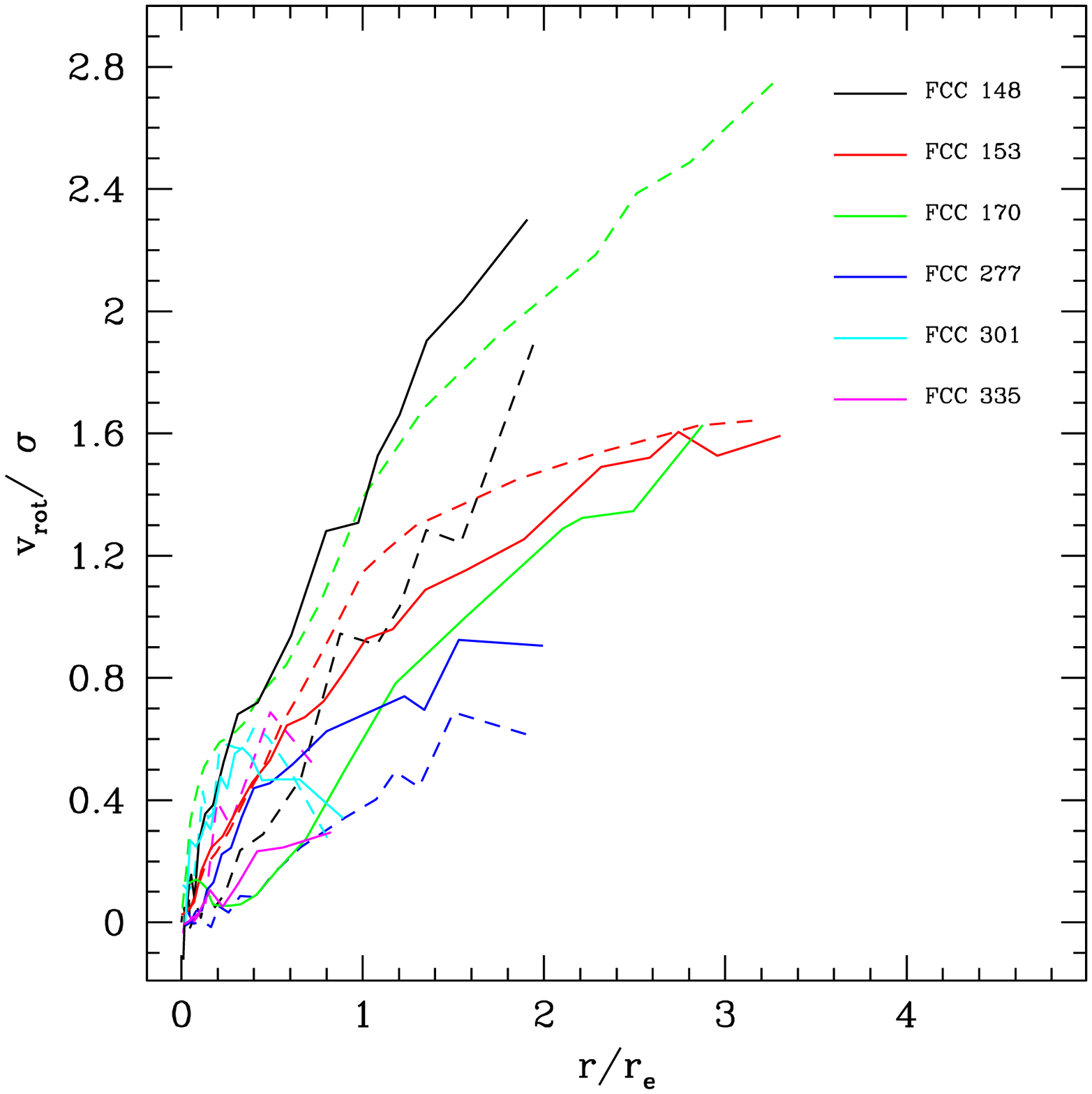}
\caption[]{Radial profiles of the rotational support parameter (ratio of local rotation velocity to local velocity dispersion) for the six Fornax cluster galaxies. The quantity is expressed as a function of the ratio between the local radius and the galaxy's effective radius $r/r_{e}$. Continuous and dashed lines indicate the two sides with respect to the galaxy centre.}
\label{fig3F}
\end{figure}

\begin{figure}
\includegraphics[scale=.4]{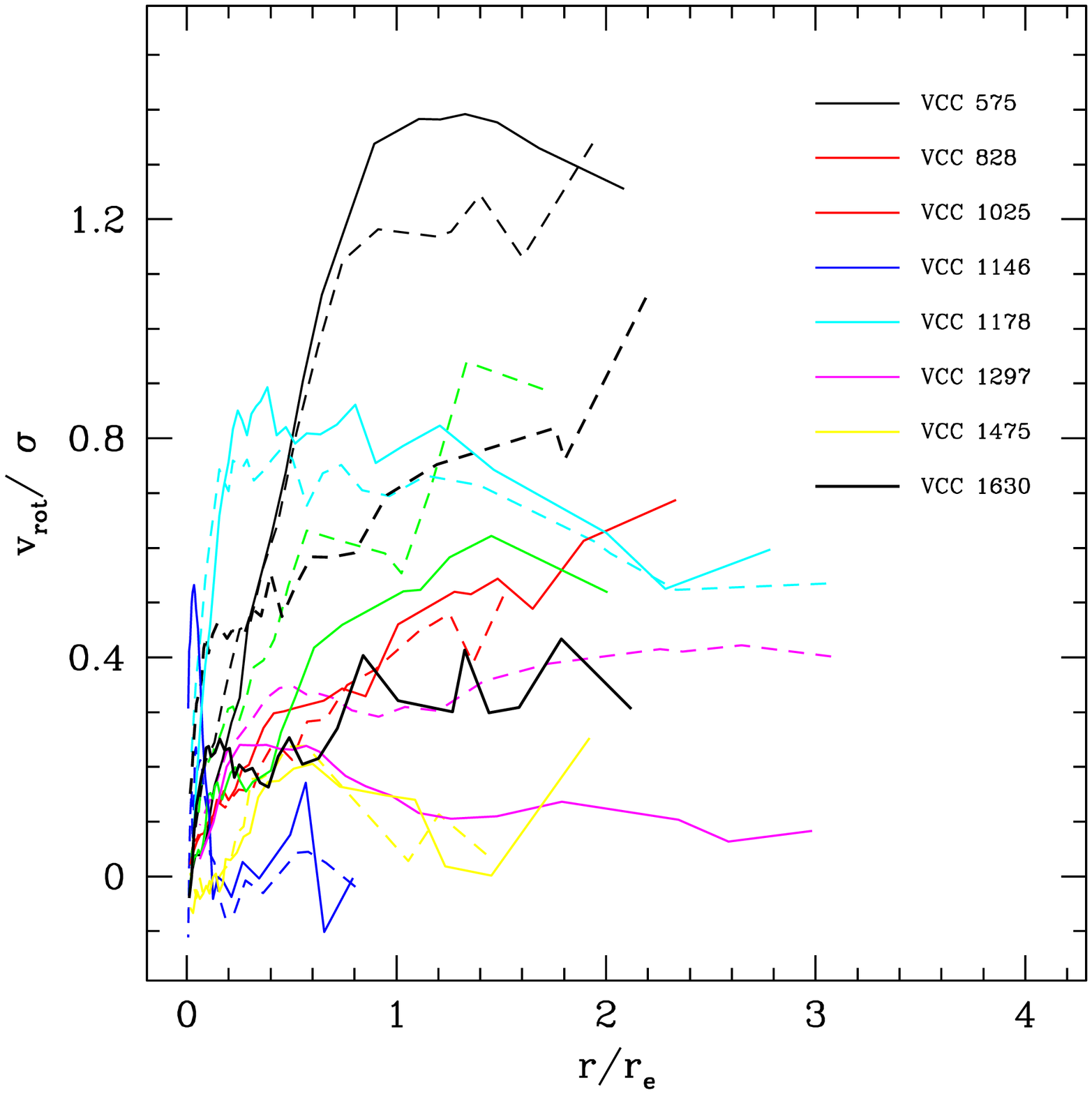}
\caption[]{Radial profiles of the rotational support parameter (ratio of local rotation velocity to local velocity dispersion) for the eight Virgo cluster galaxies. The quantity is expressed as a function of the ratio between the local radius and the galaxy's effective radius $r/r_{e}$. Continuous and dashed lines indicate the two sides with respect to the galaxy centre.}
\label{fig3V}
\end{figure}

The radial profiles of the ratio between the local rotation velocity and local velocity dispersion for our 14 low-luminosity galaxies are shown in Fig.~\ref{fig3F}, and Fig.~\ref{fig3V}.
This ratio allows us to assess the level of ordered versus random motion as a function of galactocentric radius.
The profiles have been folded onto the positive side of the major axis, with a sign change for the rotation velocity. 

For each galaxy, we find that the profiles of the two major axis's sides are similar within the errors. This suggests an overall symmetry between the two sides, and thus a large-scale  kinematic coherence.

In general, the profiles raise with the distance to centre of galaxy.
Inside one effective radius the value of $v_{rot}/ \sigma$ ratio covers the range 0 - 1.4.
At two effective radii the galaxies have $v/ \sigma$ ratio values in the range 0.6 - 2.4.
This is attributed to the rotation velocity increasing with radius and contemporarily the velocity dispersion decreasing.
In other words, the galaxy kinematics is primarily driven by the ordered motion of stars.

\subsection{Rotational support \textit{versus} Anisotropy}
\begin{figure}
\includegraphics[scale=.40]{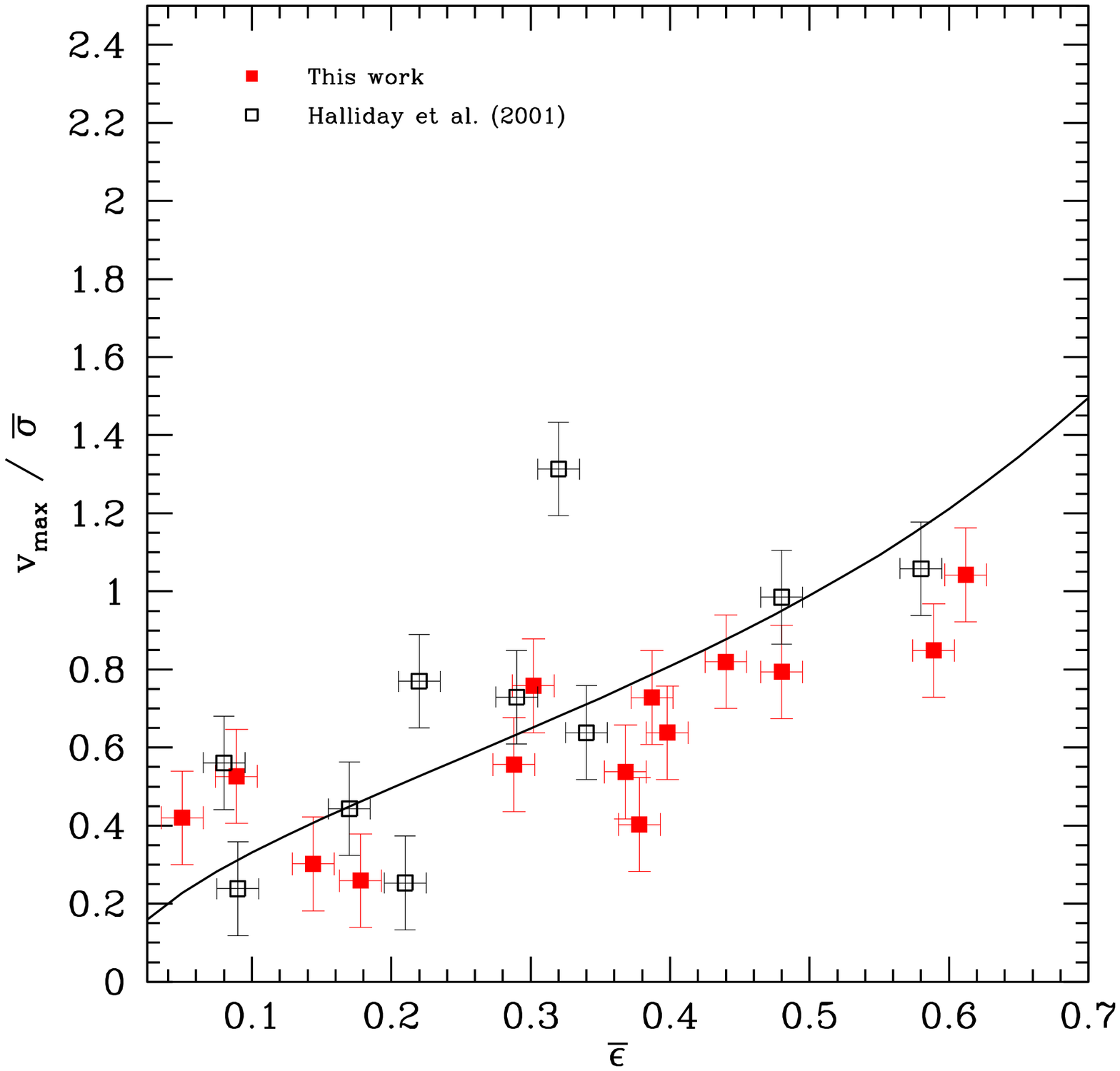}
\caption[]{Ratio of maximum rotation velocity $v_{max}$ to mean velocity dispersion $\bar{\sigma}$ plotted against mean isophotal ellipticity $\bar{\epsilon}$. Filled red squares are the data for the 14 low-luminosity galaxies presented in this paper. Open black squares are values by \cite{halliday01} for low-luminosity elliptical galaxies. The solid black line represents the expected relationship for an oblate, isotropic model galaxy flattened by rotation from \cite{binney78}.}
\label{fig1}
\end{figure}

We present the $v_{max}/ \bar{\sigma} - \bar{\epsilon}$ diagram (\citealt{binney78}) for our 14 galaxies in Fig.~\ref{fig1}.
The plot allows us to quantify the amount of rotational support of these objects with respect to their intrinsic shape.
We adopt as variables the ratio of $v_{max}$ to the mean stellar velocity dispersion 
$\bar{\sigma}$ and the mean isophotal ellipticity $\bar{\epsilon}$.

In the diagram we also show the relationship for oblate isotropic rotator models from \cite{binney78}.
The model predictions are derived from the tensor virial theorem, assuming that elliptical galaxies are oblate spheroids of constant ellipticity and with isotropic velocity distributions. The values measured for our galaxies are generally consistent with the theoretical relationship, thus being oblate rotationally flattened objects. In the plot we also show the data for low-luminosity elliptical galaxies in the Virgo cluster and Leo group from \cite{halliday01}. Four of their 14 galaxies are in common with our data sample. Their measurements are in good agreement with ours, however, we decide to use our values. 

\begin{figure}
\includegraphics[scale=.40]{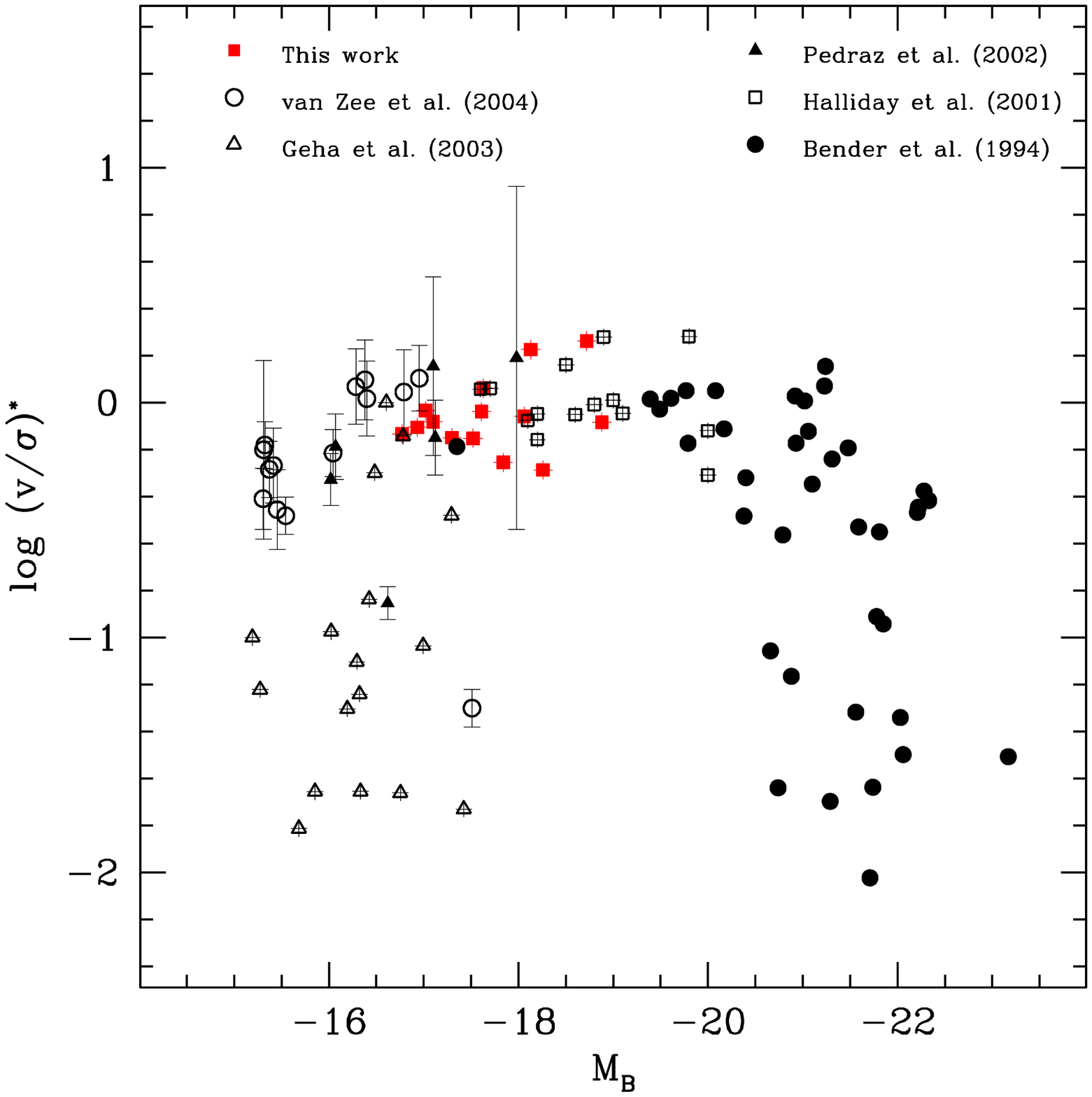}
\caption[]{Logarithm of the anisotropy parameter $(v/ \sigma)^{*}$ (ratio of the observed value of $v_{max}/ \bar{\sigma}$ to the value predicted by the isotropic model) plotted against the total $B$-band absolute magnitude. Filled red squares are the data for the 14 low-luminosity galaxies presented in this paper. Open black squares are values by \cite{halliday01} for low-luminosity elliptical galaxies. Open black circles, open black triangles and filled black triangles are values for dwarf elliptical galaxies respectively from \cite{zee04}, \cite{geha03} and \cite{pedraz02}. Filled black circles are values for bright elliptical galaxies by \cite{bender94}. }
\label{fig2}
\end{figure}

In Fig.~\ref{fig2}, we plot the anisotropy parameter $(v/ \sigma)^{*}$ against the total $B$-band absolute magnitude for each galaxy. The anisotropy parameter expresses the ratio between the observed value of $v_{max}/ \bar{\sigma}$ and the $(v/\sigma)_{OI}$ value expected for an oblate rotator (\citealt{binney78}; \citealt{davies83}). We define $(v/ \sigma)^{*}$ in terms of observable quantities as 
$(v_{max}/ \bar{\sigma})/ \sqrt{\bar{\epsilon}/(1-\bar{\epsilon})}$ (e.g.~\citealt{kormendy82}; \citealt{bender90}). 
We also plotted the data for dwarf elliptical galaxies from \cite{pedraz02}, \cite{geha03} and \cite{zee04}, low-luminosity ellipticals in \cite{halliday01} and high-luminosity elliptical galaxies from \cite{bender94}.

The anisotropy parameter is independent of ellipticity and projection effects, thus it can be used to separate galaxies with different dynamical support (e.g., \citealt{davies83}; \citealt{scorza95}).
Objects with $(v/ \sigma)^{*} < 0.7$ are anisotropically supported systems, whereas values of $(v/ \sigma)^{*} \geq 0.7$ define objects primarily supported by rotation.
Despite the scatter, the results suggest that high-luminosity elliptical galaxies are predominantly anisotropic objects and the amount of rotational support becomes increasingly significant towards intermediate and low-luminosity galaxies. The rotational component of the latter is significantly strong such that their observed flattened morphology is attributed to fast rotation. Dwarf elliptical galaxies tend to deviate from this trend, although exhibiting a range of rotational properties. 

\subsection{Rotational support \textit{versus} Discyness/Boxiness }

\begin{figure}
\includegraphics[scale=.40]{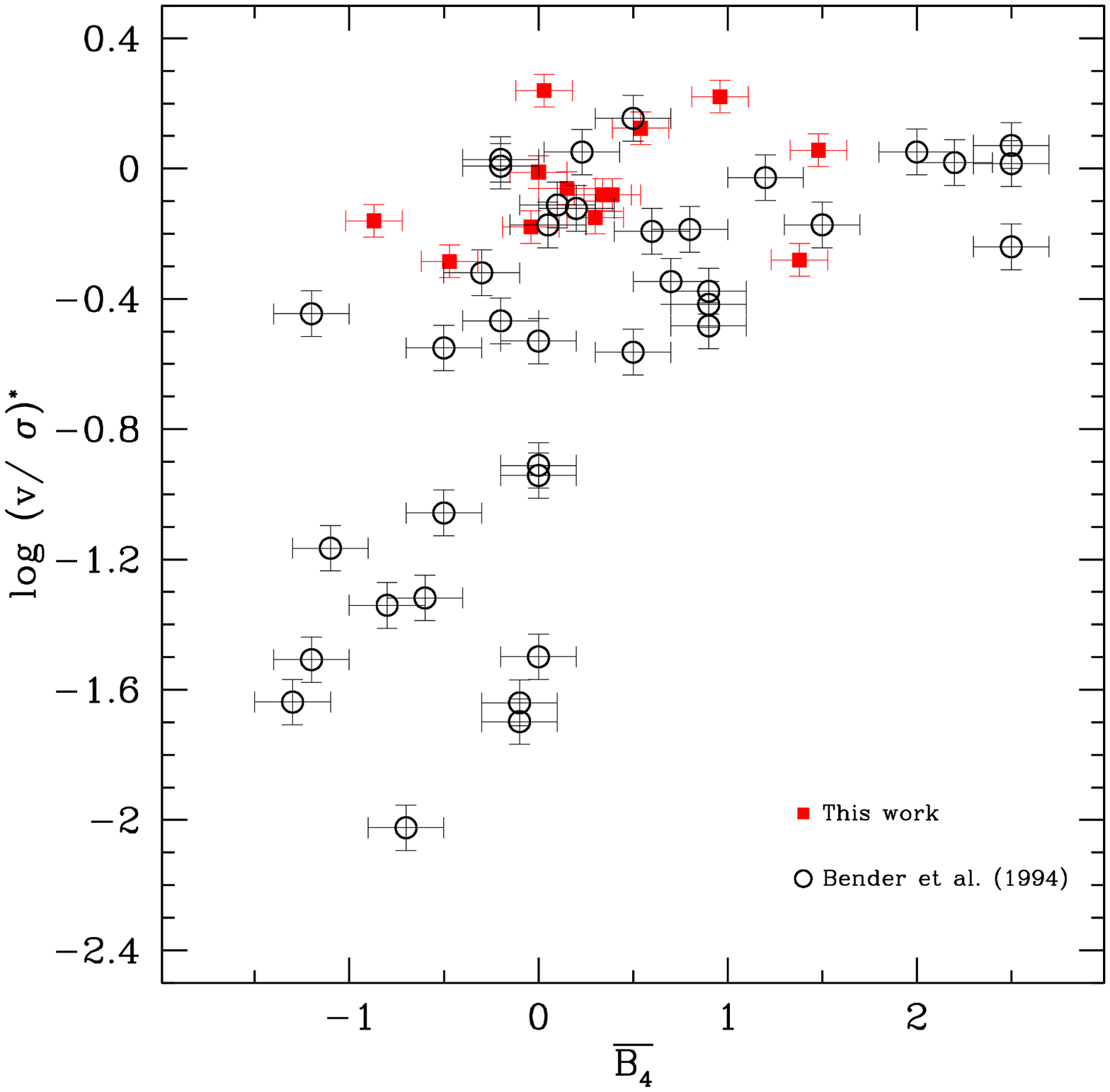}
\caption[]{Anisotropy parameter against the $\overline{B_{4}}$ coefficient. Filled red squares are the data for the 14 low-luminosity galaxies presented in this paper. Open black circles are values by \cite{bender94} for bright elliptical galaxies. Positive values of $B_{4}$ indicate discy isophotes, whereas negative values represent boxy isophotes.}
\label{fig5}
\end{figure}

In Fig.~\ref{fig5}, the relation between amount of rotational support and isophotal shape is examined.
We plot anisotropy parameter against the $\overline{B_{4}}$ coefficient of our sample galaxies, and the values of bright ellipticals by \cite{bender94} for comparison. 

There exists a relation such that galaxies with discy isophotes tend to be more rotationally supported than boxy galaxies.
The relation is independent of galaxy's luminosity (e.g., \citealt{bender94}).
We find that almost 80 percent of the low-luminosity galaxies have discy-shaped isophotes or in other terms an excess of light along the galaxy's major and/or minor axis.
The remaining objects have boxy isophotes, for which the excess of light lies along a line at 45$^{\circ}$ with respect to the galaxy's axes.

\subsection{LOSVD deviations \textit{versus} Rotational support}
\begin{figure}
\includegraphics[scale=.40]{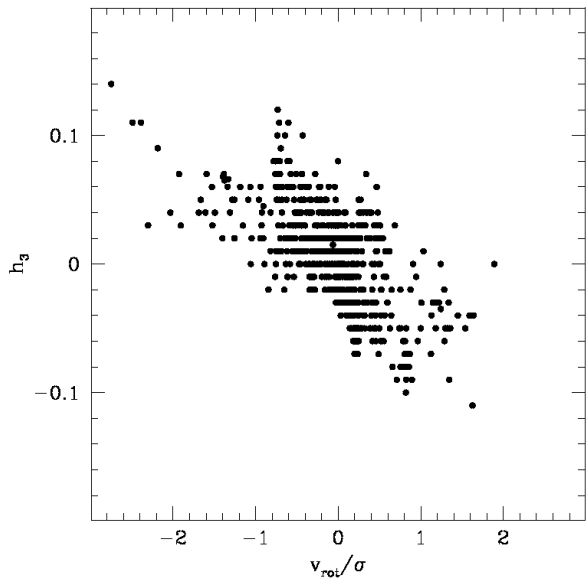}
\caption[]{Local relation between the Gauss-Hermite coefficient $h_{3}$ and the rotational parameter $v/\sigma$. Data points are from the 14 low-luminosity galaxies presented in this paper.}
\label{fig7}
\end{figure}

The plot of the skewness coefficient $h_{3}$ against the rotational support parameter $v_{rot}/\sigma$ is shown in Fig.~\ref{fig7}.
Each point corresponds to a local position along the galactocentric radius, and data for all 14 low-luminosity galaxies are presented in the same diagram.

The two parameters define a negative correlation, often referred as the $h_{3} - v_{rot}/\sigma$ plane (e.g. \citealt{bender94}).
In particular, in the central regions of galaxies, which corresponds to low $|v_{rot}/\sigma|$ values, the coefficient $|h_{3}|$ increases rapidly with $|v_{rot}/\sigma|$. 
At high values of $|v_{rot}/\sigma|$, characteristic of intermediate and large radii, $|h_{3}|$ reaches its maximum values, up to $|h_{3}| \sim 0.15$.
The relation was firstly observed by \cite{bender94}, which noted that its slope is too steep to be explained by two-integral oblate rotators models (e.g., \citealt{dehnen94}).
Instead, they suggested that ellipticals harbouring large-scale stellar discs ($\sim$ 15 percent in mass, scalength $\sim$ $r_{e}$) could explain the $h_{3} - v/\sigma$ plane.
Our results for low-luminosity galaxies confirm the validity at large radii $r > 1r_{e}$ of the $h_{3} - v_{rot}/\sigma$ distribution observed by \cite{bender94}.

\begin{figure}
\includegraphics[scale=.40]{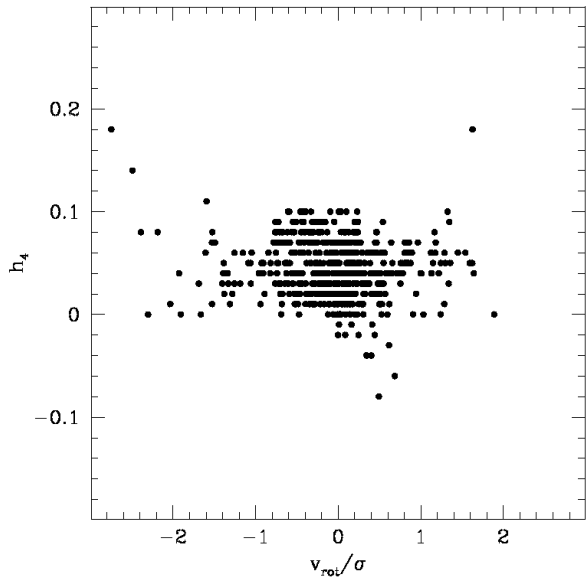}
\caption[]{Local relation between the Gauss-Hermite coefficient $h_{4}$ and the rotational parameter $v/\sigma$. Data points are from the 14 low-luminosity galaxies presented in this paper.}
\label{fig9}
\end{figure}

We now turn to the kurtosis parameter $h_{4}$, estimating the symmetric deviations of the LOSVDs from a Gaussian shape.
In Fig.~\ref{fig9} we present the $h_{4} - v_{rot}/\sigma$ distribution for all the low-luminosity galaxies.
In contrast to $h_{3}$, $h_{4}$ is not observed to correlate with the rotational parameter.
This lack of correlation was already found by \cite{bender94}. 
However, the distribution present a wide spread of positive $h_{4}$ values at any $v_{rot}/\sigma$, indicating a significant rotationally-supported disc component.

\section{Discussion}
We find that almost 85 per cent of our sample galaxies are characterised by a cold stellar component extending out to the outermost regions reached by our study, i.e.~$\sim 3 r_{e}$. The stellar velocity dispersion is observed to quickly decline with radius, such that ordered motion prevails over random stellar orbits. The galaxies are well described by oblate rotationally supported models. The results of our analysis confirms previous circumstantial findings by \cite{bender90}, \cite{davies83}, and \cite{halliday01}.
However, unlike these earlier studies, we are able to spatially resolved enough apertures at large galactocentric radii (i.e., $r \geq 0.5 r_{e}$), where the data have a sufficiently high S/N to make robust measurements.

During an early star-forming collapse (e.g., \citealt{chiosi02}; \citealt{kawata03}), a cold and extended stellar component can naturally originate from the high angular momentum gas situated in the outer regions of the forming galaxies (e.g., \citealt{bekki98}; \citealt{rix99}; \citealt{naab01}; \citealt{kawata03}). On the other hand, the gas located at inner radii tend to sink toward the galactic centre where it forms new stars.
This outside-in formation scenario (e.g., \citealt{pipino06b}) is predicted to induce a shallow positive age gradient in the galactic stellar population.
In Paper II we find that the 14 low-luminosity galaxies have an average age gradient of $0.06 \pm 0.13$ dex per decade.
The gradients are such that central regions (i.e., $r \leq 0.5r_{e}$) are a few Gyr younger with respect to the outer parts.
The stellar age of these outer galactic regions is on average older than 8 Gyr (see table~3 in Paper II)
In other words, the cold stellar component of the sample galaxies is old and it was formed at high redshift, i.e.~$z \geq1$.

The radial variation in the ellipticity of the isophotes, and their prevalent discyness, observed in our data sample could be attributed to the central concentration of stars produced by the sinking gas (e.g., \citealt{mihos96}; \citealt{naab01}). This mass congregation makes the gravitational potential well of the galaxy steeper, thus prompting a change of the stellar orbits.
Specifically, it becomes more axisymmetric and less triaxial. The lack of support for box orbits makes the minor-axis tubes more populated, thus producing discy orbits.

The properties of the LOSVDs are also affected by the presence of gas (e.g., \citealt{naab06}; \citealt{garcia06}). 
The local correlations between $h_{3,4}$ and $v_{rot}/\sigma$ observed for our low-luminosity galaxies, is reproduced in numerical simulations by superimposing a cold stellar disc component in the outer regions (i.e., $\geq 1 r_{e}$) of the spheroidal body (\citealt{naab01}). The transfer of angular momentum between stars and gas (e.g., \citealt{mihos96}) and the redistribution of angular momentum from larger radii to inner radii produce a net increase of the rotational velocity of the galaxy. 

The efficiency of gas consumption is also found to be responsible in affecting the shape of the isophotes (e.g., \citealt{bekki97}).
Discy isophotes are more likely formed in galaxies with gradual star formation, whereas galaxies with a rapid starburst tend to have boxy isophotes.
The almost solar value of the [$\alpha$/Fe] abundance ratios displayed by our galaxies (Paper II) favours this idea.
The observed abundance values are a consequence of the low star formation efficiency in our galaxies, thus leading to an extended star-forming episode ($\geq 1$ Gyr).

In conclusion, the kinematic and photometric results of our analysis are in good agreement with the interpretation of the stellar population findings in Paper II.
They suggest that an early star-forming collapse with a mass-dependent star-formation efficiency is the main mechanisms acting in the formation of these low-luminosity early-type galaxies.

Galaxy mergers may also produce some of the observed features. We now compare our results to merger simulations by \cite{naab03}, \cite{naab06} and \cite{cox06}.

\cite{naab03} used a large set of collisionless $N$-body simulations to investigate the fundamental statistical properties of merger remnants. 
Simulations involve dissipationless binary mergers of disc galaxies with mass ratios of 1:1, 2:1, 3:1, 4:1.
Projection effects are taken into account and the bulge components of the progenitor discs are left to evolve until they coalesce before the kinematic and photometric properties of the simulated early-type galaxies are measured.

\cite{naab06} re-simulated the merger remnants of \cite{naab03} with an additional gas component in the progenitor disc galaxies.
They considered only mergers of galaxies with mass ratios of 1:1 and 3:1. 
Each progenitor has 10 percent of the total mass of the stellar disc replaced by gas, but substantial heating processes prevent gas cooling and hence star formation. 
The gas is allowed to set into an equilibrium state before the disc galaxies are merged.

The study of \cite{cox06} considers mergers of equal-mass disc galaxies. 
Two sets of numerical simulations are generated to compare the statistical properties of early-type galaxies formed by dissipationless and dissipational mergers. Progenitor galaxies can contain a purely exponential stellar disc with a null fraction of gas, or disc where 40 percent of the total mass is in the form of gas. Gas-rich mergers include radiative cooling of gas, star formation, feedback processes and induced growth of massive central black holes. The orbits of the merging progenitors are identical to those explored by \cite{naab03} and \cite{naab06}, allowing a direct comparison of results.

\begin{figure}
\includegraphics[scale=.40]{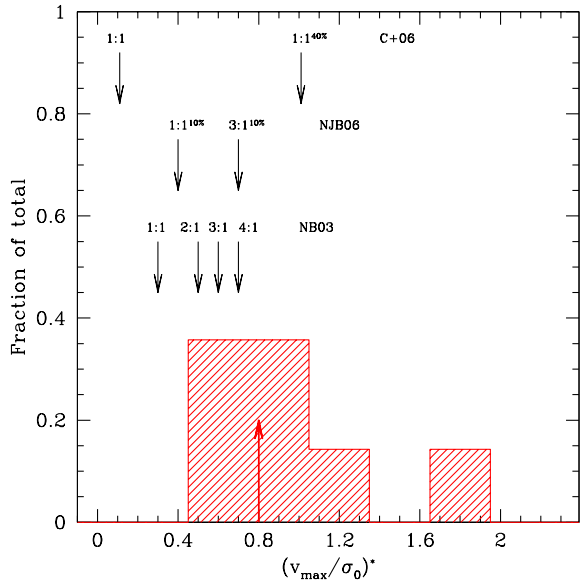}
\caption[]{Frequency distributions of the anisotropy parameter for the sample of low-luminosity early-type galaxies presented in this work (shaded red region). The median value of the distribution is shown as superimposed red arrow at the bottom of the histogram.
The three series of arrows above the histograms indicate the peaks of the distributions for simulated galaxies.
Bottom set of arrows: simulations from \cite{naab03} (labelled as NB03) for gas-free mergers with mass ratios of 1:1, 2:1, 3:1 and 4:1.
Middle set of arrows: distributions peaks from \cite{naab06} (labelled as NJB06) for mergers with mass ratios of 1:1 and 3:1 and progenitors with 10 percent of the total mass of the stellar disc replaced by gas.
Upper set of arrows: predictions from \cite{cox06} (labelled as C+06) for dissipationless and dissipational equal-mass mergers; in gas-rich mergers, progenitors have 
40 percent of the total mass of the stellar disc replaced by gas.}
\label{histo1b}
\end{figure}

In Fig.~\ref{histo1b}, we show the distribution of the anisotropy parameter of our sample in comparison with the predictions of \cite{naab03}, \cite{naab06} and \cite{cox06}.
The distribution of our low-luminosity galaxies has a median value of $(v_{max}/\sigma_{0})^{*}$$\sim 0.86$ and the values span the range 0.4 - 2.0.
The distributions of simulated remnants have significantly different median values for the degree of rotational support with respect to our galaxy sample.
However, there exists a tendency for mergers with increasingly different mass ratios to produce more isotropically supported galaxies.
The predictions of \cite{naab06} and \cite{cox06} show that an additional gas component in the progenitor discs has a significant impact on the galaxy dynamical support.
This dissipative component increase the amount of rotation and therefore ordered motion in the remnants.
The results of a Kolmogorov$-$Smirnov (KS) test show that our observed galaxies and galaxies simulated by \cite{cox06} have a difference in their anisotropy parameter distributions 
significant at 4$\sigma$ level. On the other hand,  remnants of gas-free mergers with mass ratios of 4:1 simulated by \cite{naab03} and merger models of \cite{naab06} with mass ratios 3:1 with progenitors having 10 percent of gas appear similarly distributed to our sample galaxies at almost the 60 per cent level.

\begin{figure}
\includegraphics[scale=.40]{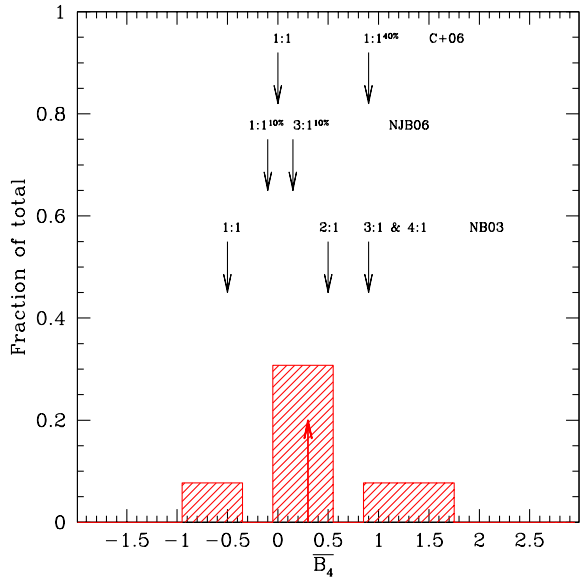}
\caption[]{Frequency distributions of the coefficient $\overline{B_{4}}$ for the sample of low-luminosity early-type galaxies presented in this work (shaded red region). Description is as in Fig.~\ref{histo1b}.}
\label{histo3b}
\end{figure}

In Fig.~\ref{histo3b}, the observed distribution of the isophote-shape parameter $\overline{B_{4}}$ (the discyness/boxiness parameter) is compared to merger predictions.
The sample galaxies have discy-shaped isophotes, with a distribution median value of $\overline{B_{4}} =$ 0.30.   
The dissipationless models of \cite{naab03} produce merger remnants with more discy isophotes when progenitors of increasingly different mass ratio are considered.
On the other hand, the simulations of \cite{naab06} suggest that the presence of a gas component in the progenitors is responsible for changing the stellar orbits of the remnant.
In this case, progenitors with a mass ratio 3:1 and 10 percent of a gas component produce remnants with a small amount of discy isophotes.
A KS test marginally rejects at the 70 per cent level the hypothesis that dissipationless merger with mass ratios of 2:1 and 10 per cent gas-rich mergers with mass ratios 3:1 
have the same distribution of our sample galaxies.
The simulations of \cite{cox06} show that equal-mass progenitors with 40 percent of gas and merger-induced star formation produce remnants with significantly discy isophotes. The KS test confirms that the degree of discyness of the simulated galaxies is inconsistent with that one of the observed galaxies.
We note that the inclination angle under which the galaxy is observed might affect the evaluation of the isophotes's shape (i.e. projection effects).

In conclusion, the kinematic and isophotal shape predictions from merger models examined are in less good agreement with our results.
Although we suggest that merging is unlikely to be the main galaxy formation mechanism of our galaxies, comparison with the model predictions provides us with some constraints on the merger properties.
In order to reproduce the observed properties the mergers are required to happen at high-redshift ($z \geq 1$), between progenitors of different mass ratio (at least 3:1) and with a significant fraction of the total mass in the form of gas (i.e., $\geq 10$ percent). 
However, this formation scenario is not able to recreate the the observed mass-metallicity gradient relation of these low-luminosity galaxies (\citealt{spolaor09a}; Paper II).

A further interpretation of our results is that low-luminosity galaxies were originally late-type galaxies, whose star formation has been truncated by removal of gas (i.e., strangulation) and subsequently the disc has been dynamically heated by high velocity encounters (i.e., galaxy harassment) in the cluster environment. Simulations have shown that late-type galaxies entering in a rich cluster can undergo a significant morphological transformation into spheroidals by encounters with brighter galaxies and the cluster's tidal field (\citealt{moore96}; \citealt{mastropietro05}). In this scenario, we expect the original disc to be dynamically heated by the interactions such that stellar orbits acquire a significant velocity component perpendicular to the disc. The imprint of their previous morphological nature is preserved in the form of an embedded stellar disc (e.g., \citealt{rijcke03}; \citealt{chilingarian08}). For example, \cite{beasley09} reported significant rotation at large radii (i.e., $4 -7 r_{e}$) of two luminous Virgo dwarf ellipticals, using globular cluster systems as tracers of galaxy dynamics. They show that the detection of such large amount of rotation in the outer galactic regions support the idea that luminous dwarf ellipticals were originally disc galaxies. 
Numerical simulations (\citealt{mastropietro05}) predict values of the anisotropy parameter $(v/\sigma)^{*}$ similar to those of our galaxies, and also disky isophotes with similar $\overline{B_{4}}$ values. Moreover, the lack of counter rotation and the high incidence of coupled rotation between disc and bulge observed in our sample galaxies may favour a disc-heating scenario, whereby stars of the original disc contribute towards the bulge population while retaining some of the angular momentum.

However, this scenario is more difficult to reconcile with the stellar population properties, and in particular the mass-metallicity gradient relation, observed in our galaxies. We expect the star formation to be truncated in the disc due to the interactions. Thus, metallicity gradients are required to form in the late-type galaxies and to be somehow preserved during the high velocity encounters. We conclude that although the kinematic and isophotal features in our galaxy sample can be interpreted in the context of a morphological transformation from late to early types, further detailed numerical simulations are needed to understand if such a scenario can also explain the stellar population trends observed.

\section{Conclusions}
We have investigated the kinematic and photometric properties at large galactocentric radii for a sample of 14 low-luminosity early-type galaxies.
The radial extent considered in our analysis, i.e. $\sim 1 - 3 r_{e}$, allows us to probe more than 50 percent of the total baryonic mass of a galaxy and more than $\sim$ 10 percent of its angular momentum. We derive radial profiles for the kinematic parameters of rotational velocity, velocity dispersion, and the Gauss-Hermite coefficients $h_{3}$ and $h_{4}$. We also obtain radial profiles for the photometric parameters in both $g$ and $z$-band of surface brightness, ellipticity, position angle and discyness/boxyness.

We find that our low-luminosity early-type galaxies host an extended, cold, and old (i.e., $\geq 8$ Gyr) stellar component at radii larger than $0.5 r_{e}$. This reaches out to the largest radii (i.e., $\sim 3 r_{e}$) and the amount of rotational support increases as a function of the galactocentric radius. This implies that the low-luminosity galaxies are predominantly supported by rotation. The isophotes are discy-shaped and their degree of ellipticity is such that these galaxies can be well described by oblate isotropic rotator models. 

The observed properties suggest that an early star-forming collapse is the main mechanism acting in the formation of these galaxies.
The old and extended cold stellar component is generated from the initial high angular momentum gas in the outer regions of the forming galaxy. 
The low star formation efficiency of these galaxies, endorsed by their stellar population properties (Paper II), favours the observed isophotal discyness and the isophotes' ellipticity radial variation.
Our results support and complement the interpretation of the stellar populations findings in Paper II.

Major mergers are unlikely to be the main mechanisms responsible for the formation of these galaxies.
In particular, numerical simulations of different merger categories fail to reproduce the amount of rotational support and isophotal shape observed in the low-luminosity galaxies. Although we can not rule out a minor merging origin for these galaxies, the comparison of our results with theoretical predictions provides us with some constraints on the possible merger progenitors. In order to reproduce the observed properties, the mergers are required to happen at high-redshift (i.e., $z \geq 1$) involving progenitors of different mass ratio (at least 3:1) and with a significant amount of gas content (i.e., $\geq$ 10 percent).

An alternative scenario is that the low-luminosity galaxies were formerly late-type galaxies, whose star formation has been truncated by removal of gas and subsequently the disc has been dynamically heated by high speed encounters in the cluster environment. However, we conclude that further detailed numerical simulations are needed to understand if such disc-heating scenario can also reproduce the stellar population properties of our sample galaxies. 

\section{Acknowledgement}
Based on observations obtained at the Gemini Observatory, which is operated by the
Association of Universities for Research in Astronomy, Inc., under a cooperative agreement
with the NSF on behalf of the Gemini partnership: the National Science Foundation (United
States), the Science and Technology Facilities Council (United Kingdom), the
National Research Council (Canada), CONICYT (Chile), the Australian Research Council
(Australia), Minist\'{e}rio da Ci\^{e}ncia e Tecnologia (Brazil) 
and Ministerio de Ciencia, Tecnolog\'{i}a e Innovaci\'{o}n Productiva  (Argentina).

Some of the data presented in this paper were obtained from the Multimission Archive at the Space Telescope Science Institute (MAST). STScI is operated by the Association of Universities for Research in Astronomy, Inc., under NASA contract NAS5-26555. Support for MAST for non-HST data is provided by the NASA Office of Space Science via grant NAG5-7584 and by other grants and contracts. We made used of IRAF, that is distributed by the National Optical Astronomy Observatories, which are operated by the Association of Universities for Research in Astronomy, Inc., under cooperative agreement with the National Science Foundation.

We thank Andres Jordan (P.I.) and Laura Ferrarese for providing photometric parameters of the Fornax cluster galaxies in advance of publication. MS thanks the Centre for Astrophysics \& Supercomputing for providing travel support.
We thank S. Burke-Spolaor for a careful reading of the manuscript.
We thank Henry Lee and the other staff astronomers at the Gemini South Telescope for the support given to our observing programs (GS-2006B-Q-74; GS-2008A-Q-3).
We also thank the anonymous referee for his/her constructive comments.

\begin{appendix}
\def\ang{\mbox{\AA}}
\def\kms{\mbox{${\rm km\ s}^{-1}$}}
\def\ergsec{\mbox{${\rm erg}\thinspace{\rm s}^{-1}$}}
\def\ergsqcmsec{\mbox{${\rm erg \thinspace cm}^{-2}{\rm s}^{-1}$}}
\def\deg{\hbox{$^\circ$}}
\def\min{\hbox{$^{'}$}}
\def\asec{\hbox{$^{"}$}}
\def\msun{\mbox{${\rm M}_{\odot}$}}
\def\rsun{\mbox{${\rm R}_{\odot}$}}
\def\gsun{\mbox{${\rm g}_{\odot}$}}
\def\lsun{\mbox{${\rm L}_{\odot}$}}
\def\msunyr {\mbox{M$_{\odot}\mbox{yr}^{-1}$}}
\def\etal{et~al.\ }

\section{Individual galaxies}
\label{ind_gals}
\subsection{FCC 148 (NGC 1375)}
The rotation of FCC 148 increases outward to a maximum amplitude of $\sim 89\ \kms$, where $\sigma$ falls to $\sim 40\ \kms$. The $h_3$ has the opposite sign to the rotation, often seen in systems with an embedded cold disk. 

The increase in rotational support outwards is also reflected in the increasingly elliptical isophotes with radius, with the ellipticity rising from $\sim 0.2$ near the galaxy centre to $> 0.6$ at $r  >  1 r_e$ (1.3 kpc). 

\subsection{FCC 153 (IC 1963)}
The rotation of FCC 153 increases outward and plateaus at  $\sim 133\ \kms$, where $\sigma$ falls to $\sim 30\ \kms$. The $\sigma$ profile is slightly asymmetric in the centre, being higher on the negative $r$ side.
The $h_3$ has the opposite sign to the rotation, often seen in systems with an embedded cold disk. 

Like FCC 148 the increase in rotational support outwards is also reflected in the increasingly elliptical isophotes with radius, with the ellipticity rising from $\sim 0.2$ near the galaxy centre to $\approx 0.6$ at $r  >  0.2 r_e$ (220 pc). The isophotes between $0.1$ and $0.4$ $r_e$ (110 pc and 440 pc) are also significantly disky. The surface brightness $\mu$ only slightly decreases with radius in the galaxy centre, and exhibits an abrupt break at $r \sim 0.9 r_e$, roughly at the transitional point where the rotation plateaus. Thus the correlation between the surface brightness, rotation and ellipticity may be caused by a radially varying disk-to-bulge ratio.

\subsection{FCC 170 (NGC 1381)} 
The kinematics of FCC 170 exhibits several peculiarities. Firstly, over the monotonically outwardly increasing rotation, the rotation is enhanced in the central regions ($|r| < 0.3\ r_e$), where $\sigma$ is also peaked, and the $h_3$ has an opposite sign to the rotation. This seems to suggest there is a small nuclear embedded disk with radius $\approx 370$ pc. In the outer parts of the galaxy the rotation is strong, at $\sim 200\ \kms$, where the $h_3$ has an opposite sign to the rotation. This suggests substantially skewed LOSVDs at large radii, which could be the signature of an extended disk. This seems to be supported by the photometry, which shows that the isophotes are increasingly elliptical with increasing $r$. The ellipticity rises from $\sim 0.2$ at $0.1 r_e$ (120 pc) to 0.5 at 1 $r_e$ (1.2 kpc). 

\subsection{FCC 277 (NGC 1428)}
The kinematics of FCC 277 bears some resemblance to that of FCC 148, with an outwardly rising rotation which reaches $\sim 68$ \kms. There is also a weak asymmetry in the LOSVD indicated by anti-correlation between $h_3$ and rotation.

The increase in rotational support outwards is also reflected in the increasingly elliptical isophotes with radius, with the ellipticity rising from $\sim 0.1$ at $r \sim 0.1 r_e$ (90 pc) to $> 0.45$ at $r  >  2 r_e$ (1.8 kpc). 
 
\subsection{FCC 301 (ESO 358-G059)}

The most notable feature of FCC 301 is an outwardly rising $\sigma$ profile and a depressed central $\sigma$, which is the reverse of most other galaxies. The rotation seems to peak at $|r| \sim 0.3 r_e$ with amplitude of $\sim 50$ \kms and then falls off. It is tempting to interpret the depressed central $\sigma$ as due to the presence of a luminous, dynamically cold sub-component of radius $~0.3 r_e$ ($289$ pc) in the galaxy centre. Farther out, the higher $\sigma$ ($\sim 80$ \kms) and the lower rotation indicate an outward increase in dispersion support.

The photometry supports the hypothesis of a central kinematic subcomponent. The surface brightness is flat or slightly decreasing with radius within $0.3 r_e$ (280 pc) of the galaxy centre, and falls off beyond that. The isophotes has a dramatic rise in ellipticity within the same region, reaching $\epsilon \sim 0.6$, and then falls with distance beyond that.

\subsection{FCC 335 (ESO 359-G002)}
The rotation of FCC 335 is not symmetric. For positive $r$ the stars' mean rotation gently rises outwards to $\sim 15$ \kms, while for negative $r$ the stellar mean rotational amplitude reaches a maximum at $\sim 30$ at $0.5 r_e$ (700 pc). There is also asymmetry in $\sigma$. In addition, the $\sigma$ is centrally peaked inside $\sim 0.1 r_e$ (140 kpc), where the $h_4$ is also peaked. The galaxy has an outwardly rising ellipticity, which reaches a maximum of 0.4.


\subsection{VCC 575 (NGC 4318)}
The kinematics of VCC 575 is quite similar to that of FCC 153, but with lower rotation and $\sigma$. The presence of an embedded disk is also suggested

The increase in rotational support outwards is also reflected in the increasingly elliptical isophotes with radius, with the ellipticity rising from $\sim 0.2$ at $r \sim 0.4 r_e$ to $> 0.35$ at $0.5 > r  >  2 r_e$. The isophotes are also significantly disky beyond 1 $r_e$. The surface brightness $\mu$ exhibits a break at $r \sim 0.8 r_e$, roughly at the transitional point where the rotation plateaus. Thus the correlation between the surface brightness, rotation and ellipticity may be caused by a radially varying disk-to-bulge ratio.

\subsection{VCC 828 (NGC 4387)}

The most striking feature is a distinct kinematic feature within $0.25$ $r_e$ (200 pc) of the nucleus, indicated by a centrally lowered dispersion and an enhanced velocity gradient compared to the rest of the galaxy. This suggests that embedded within the slowly rotating galaxy is a bright, dynamically cold small disk component with a higher rotational speed. The $h_3$ does not exhibit any significant feature associated with the decoupled core. This may suggest that the central disk dominates the brightness there causing only a weak deviation from a gaussian LOSVD.

The photometry also support the kinematic findings. The ellipticity profile also shows a break at the location where the dispersion changes gradient, with $\epsilon$ roughly constant at $0.2$ inside $0.2\ r_e$, and outward rising beyond that. There is also a break in the $B_4$ profile at the same location, indicating that the isophotes go from elliptical to slightly boxy from $0.2\ r_e$ outwards.
Our results are in good agreement with the previously kinematic measurements by \cite{halliday01}.

\subsection{VCC 1025 (NGC 4434)}
The kinematics of VCC 1025 shows signs of an extended embedded disk, as indicated by the asymmetric LOSVD with opposing $v$ and $h_3$. At the outerparts of the galaxy $v_{rot}$ and $\sigma$ have roughly similar amplitude, indicating comparable rotational and dispersion support. Photometrically the galaxy is fairly round thoughout. It is possible that the disk is inclined.

\subsection{VCC 1146 (NGC 4458)}
The most striking feature is the presence of a compact kinematically distinct core. The galaxy overall has little rotation and is dispersion supported. However, within $\sim0.1 r_e$ of the nucleus (210 pc), the rotation is significantly enhanced and the LOSVD is distinctly skewed, with $h_3$ having opposite sign to $V_{rot}$. Such feature can be attributed to the presence of a nuclear disk. The $\sigma$ is strongly peaked at the nucleus, and falls sharply outwards. This fall is associated with the sharp rise in rotation, due to the contribution of the disk component to the galaxy surface brightness.

The compact kinematically distinct core is also detected in the photometry. There is an abrupt break in ellipticity roughly where the aforementioned kinematic transition occurs. The nuclear region has a high ellipticity ($\epsilon \approx 0.4$) while the regions beyond $0.1 r_e$ are fairly round ($\epsilon < 0.15$). The kinematically distinct core is also slightly disky compared to the rest of the galaxy.
Our results are in good agreement with the previously kinematic measurements by \cite{halliday01}.

\subsection{VCC 1178 (NGC 4464)}
VCC 1178 appears to possess a large scale disk, as indicated by a rotation curve which resembles that of a disk galaxy, with a sharply rising rotation which reaches a maximum within $0.2 r_e$ (100 pc), and then gradually declines outwards. The high asymmetry in the LOSVD also supports this hypothesis. At $r \approx 100$ pc, where the rotation reaches its maximum, the dispersion is roughly half that of the nuclear value. The kinematics is generally consistent with the presence of a dispersion-supported bulge plus a large scale disk.

The photometry also indicates a radially changing disk-to-bugle ratio. The ellipticity profile has a peak at $0.2 r_e$, and is lower in the galaxy center and the outer parts. Interpreting the increase in ellipticity is caused an increase in light contribution from a disk component, both kinematics and photometry indicate  that the large scale disk has a maximum contribution at $0.2\ r_e$.
Our results are in good agreement with the previously kinematic measurements by \cite{halliday01}.

\subsection{VCC 1297 (NGC 4486B)}
The kinematics of VCC 1297 bears some resemblance to that of VCC 1178, with the exception that the dispersion is comparably higher throughout the galaxy, the LOSVD asymmetry is weaker, and the decline in the dispersion profile is less dramatic. This may suggest a smaller contribution to the total surface brightness from the large-scale disk, compared to VCC 1178.

The photometry shows a generally outwardly decreasing ellipticity with radius. The ellipticity has a ``shelf'' between roughly 0.1 and 0.4 $r_e$, where the rotational is fastest. As the rotation gradually declines to an asymtoptic value, the galaxy becomes rounder. These seem to suggest that the large scale disk is more dominant at the galaxy centre than in the outer parts, and VCC 1297 has a higher support from dispersion.

\subsection{VCC 1475 (NGC 4515)}
The most striking kinematic feature is the lowered dispersion within $0.2 r_e$ (150 pc) of the nucleus. Within this region there is a mild counter-rotation of $\sim 5\ $ \kms with respect to the rest of the galaxy, which has a maximum rotation of $\sim 22\ \kms$. There is significant asymmetry in the LOSVD, with $h_3$ having the opposite sign to $v_{rot}$. The $h_3$ also changes sign on opposite sides of the nucleus. If the kinematically distinct core is caused by the light dominance of a kinematically cold subcomponent, then this subcomponent should exhibit quite pronounced counter-rotation. At first sight, it seems puzzling why the LOSVD asymmetry is pronounced, as indicated by the $h_3$, while counter-rotation is so weak as indicated by $v_{rot}$. It is unlikely that the kinematic subcomponent is a non-rotating dwarf or star cluster, or a disk which is orientated along the slit.  It is possible that there exists a cold disk which is not aligned with the slit, such that little rotation is sampled along the slit, while there exists enough asymmetry in the LOSVD to give rise to the $h_3$ profile. 

The photometry seems to support the above hypothesis. The P.A. is $\sim 60$ throughout the galaxy, thus indicating that the slit is 45 degrees offset from the major axis. The "disk" is associated with a region of high ellipticity ($\epsilon \sim 0.5$ and significant diskyness ($B_4 \sim 0.03$). Farther out, the galaxy becomes progressively round and exhibit a high degree of dispersion support.

\subsection{VCC 1630 (NGC 4551)}
The galaxy's rotation resembles that of a disk galaxy, with a maximum amplitude of $\sim 70$ \kms. The galaxy lies close to the dividing line between rotational and dispersion support. A striking feature is the large asymmetry in the LOSVD within $\sim 0.3 r_e$  (300 pc) of the nucleus, indicated by the $h_3$ which has the opposite sign to $v_{rot}$. Over the same region $\sigma$ is depressed.
The kinematics are consistent with the presence of a nuclear cold disk as well as a fainter outer disk, which could be the manifestation of a single structure due to a radially varying disk-to-bulge ratio.

The photometry does not show any particular outstanding feature. The galaxy is moderately elliptical with ellipticity roughly fluctuating around $\epsilon \sim 0.3$. There is hint of a slightly enhanced ellipticity and diskyness within $r \sim 0.3 r_3$. But the overall lack of a distinctive photometric feature seems to indicate the presence of a single embeded disk which give rise to the aforementioned kinematic features.
Our results are in good agreement with the previously kinematic measurements by \cite{halliday01}.

\end{appendix}
\end{document}